\documentclass{elsarticle}

\usepackage{amsfonts,amsmath,amssymb}
\usepackage{enumerate}
\usepackage{graphicx}
\usepackage{hyperref}
\usepackage{listings}
\usepackage{url}

\newcommand{\nd}{\noindent}

\newcommand{\eqs}{\,{=}\,}
\newcommand{\ges}{\,{\ge}\,}
\newcommand{\I}{\mathcal{I}}
\newcommand{\ins}{\,{\in}\,}
\newcommand{\KB}{\mathcal{K}}
\newcommand{\les}{\,{\le}\,}
\newcommand{\mids}{\,{\mid}\,}
\newcommand{\unit}{[0,1]}

\newtheorem{example}{Example}

\journal{arXive.org}

\begin{document}

\begin{frontmatter}

\title{Fuzzy Ontology Representation using OWL 2~\tnoteref{t1}}
\tnotetext[t1]{This paper is a revised and considerably extended
version of ``Representing Fuzzy Ontologies in OWL 2'', published in
the Proceedings of the 19th IEEE International Conference on Fuzzy
Systems (FUZZ-IEEE 2010).}

\author[label1]{Fernando Bobillo}
\ead{fbobillo@unizar.es}

\author[label2]{Umberto Straccia}
\ead{straccia@isti.cnr.it}

\address[label1]{Department of Computer Science and Systems Engineering, University of Zaragoza, Spain}
\address[label2]{Istituto di Scienza e Tecnologie dell'Informazione (ISTI - CNR), Pisa, Italy}

\begin{abstract}

The need to deal with vague information in Semantic Web languages is
rising in importance and, thus, calls for a standard way to
represent such information. We may address this issue by either
extending current Semantic Web languages to cope with vagueness, or
by providing a procedure to represent such information within
current standard languages and tools. In this work, we follow the latter
approach, by identifying the syntactic differences that a fuzzy
ontology language has to cope with, and by proposing a concrete
methodology to represent fuzzy ontologies using OWL 2 annotation
properties. We also report on the prototypical implementations.

\end{abstract}


\end{frontmatter}


\section{Introduction}

\nd Today, there is a growing interest in the development of knowledge
representation formalisms able to deal with uncertainty, which is a
very common requirement in real world applications. Despite the
undisputed success of ontologies, classical ontology languages are
not appropriate to deal with vagueness or imprecision in the
knowledge, which is inherent to most of the real world application
domains~\cite{URW3}.

Since fuzzy set theory and fuzzy logic~\cite{Zadeh65} are suitable
formalisms to handle these types of knowledge, fuzzy ontologies
emerge as useful in several applications, ranging from (multimedia)
information retrieval to image interpretation, ontology mapping,
matchmaking, decision making, or the Semantic Web~\cite{FLandSW}.

Description Logics (DLs for short)~\cite{DLHandbook} are a family of
logics for representing structured knowledge. Each logic is denoted
by using a string of capital letters which identify the constructors
of the logic and therefore its complexity. DLs have proved to be
very useful as ontology languages. For instance, the language OWL 2,
which has very recently become a W3C Recommendation for ontology
representation~\cite{OWL2JWS,OWL2syntax}, is equivalent to the DL
$\mathcal{SROIQ}\mathbf{(D)}$.

Several fuzzy extensions of DLs can be found in the literature (see
the survey in~\cite{StracciaSurvey}) and some fuzzy DL reasoners
have been implemented, such as
\textsc{fuzzyDL}~\cite{BobilloFuzzIEEE2008},
\textsc{DeLorean}~\cite{BobilloURSW2008} and
\textsc{Fire}~\cite{StoilosFire}. Not surprisingly, each reasoner
uses its own fuzzy DL language for representing fuzzy ontologies
and, thus, there is a need for a standard way to represent such
information.

A first possibility would be to adopt as an standard one of the
fuzzy extensions of the languages OWL and OWL 2 that have been
proposed~\cite{FuzzyOWLChino,StoilosSROIQ,StoilosIJAR}. However, we
do not expect a fuzzy OWL extension to become a W3C proposed
standard in the near future. Furthermore, we argue that current
fuzzy extensions are not expressive enough, as they only provide
syntactic modifications in the ABox.

In this work, we propose to use OWL 2 itself to represent fuzzy
ontologies. More precisely, we use OWL 2 annotation properties to
encode fuzzy $\mathcal{SROIQ}\mathbf{(D)}$ ontologies. The use of
annotation properties makes possible \emph{(i)} to use current OWL 2
editors for fuzzy ontology representation, and \emph{(ii)} that OWL
2 reasoners discard the fuzzy part of a fuzzy ontology, producing
the same results as if would not exist. Additionally, we identify
the syntactic differences that a fuzzy ontology language has to cope
with, and show how to address them using OWL 2 annotation
properties.

The remainder of this paper is organized as follows. In
Section~\ref{sec:fuzzySROIQD} we present a fuzzy extension of DL
$\mathcal{SROIQ}\mathbf{(D)}$, the logic behind OWL 2, including
some additional constructs, peculiar to fuzzy logic.
Section~\ref{sec:owl2} discusses how to encode it using OWL 2
language. Section~\ref{sec:examples} illustrates the methodology
with some application problems. Section~\ref{sec:discussion}
discusses the implementation status of our approach and compares it
with the related work. Finally, Section~\ref{sec:conclusions} sets
out some conclusions and ideas for future research.


\section{Fuzzy Logic}
~\label{sec:fuzzyLogic}

\nd Fuzzy set theory and fuzzy logic were proposed by L.
Zadeh~\cite{Zadeh65} to manage imprecise and vague knowledge. While
in classical set theory elements either belong to a set or not, in
fuzzy set theory elements can belong to a set to some degree. More
formally, let $X$ be a set of elements called the reference set. A
\emph{fuzzy subset} $A$ of $X$ is defined by a membership function
$\mu_{A}(x)$, or simply $A(x)$, which assigns any $x \in X$ to a
value in the interval of real numbers between $0$ and $1$. As in the
classical case, $0$ means no-membership and $1$ full membership, but
now a value between $0$ and $1$ represents the extent to which $x$
can be considered as an element of $X$.

Changing the usual true/false convention leads to a new concept of
statement, whose compatibility with a given state of facts is a
matter of degree, usually called the \emph{degree of truth} of the
statement. In this article we will consider \emph{fuzzy statements}
of the form $\phi \ges \alpha$ or $\phi \les \beta$, where
$\alpha,\beta \ins \unit$~\cite{Hajek98} and $\phi$ is a statement.
This encodes the fact that the degree of truth of $\phi$ is \emph{at
least} $l$ (resp. \emph{at most} $u$). For example,
$\textsf{ripeTomato} \ges 0.9$ says that we have a rather ripe
tomato (the degree of truth of \textsf{ripeTomato} is at least
$0.9$).

All crisp set operations are extended to fuzzy sets. The
intersection, union, complement and implication set operations are
performed by a t-norm function, a t-conorm function, a negation
function and an implication function, respectively. These operations
can be grouped in families or fuzzy logics. It is well known that
different fuzzy logics have different properties~\cite{Hajek98}.

There are three main fuzzy logics: {\L}ukasiewicz, G\"{o}del, and
Product. The importance of these three fuzzy logics is due the fact
that any continuous t-norm can be obtained as a combination of
\L{}ukasiewicz, G{\"o}del, and Product
t-norm~\cite{MostertShieldsTheorem}. It is also common to consider
the fuzzy connectives originally considered by Zadeh (G{\"o}del
conjunction and disjunction, \L{}ukasiewicz negation and
Kleene-Dienes implication), which is sometimes known as Zadeh fuzzy
logic. Table~\ref{tab:fuzzyOperators} shows these four fuzzy logics:
Zadeh, {\L}ukasiewicz, G\"{o}del, and Product.

\begin{table}[htbp]
\caption{Some popular fuzzy logics} \label{tab:fuzzyOperators} \hbox
to \textwidth{ \hss {\scriptsize
\begin{tabular}{|l|l|l|l|l|}
    \hline
    Family & t-norm $\alpha \otimes \beta$ & t-conorm $\alpha \oplus \beta$ & negation $\ominus \alpha$ & implication $\alpha \Rightarrow \beta$ \\
    \hline \hline
    Zadeh & $\min \{\alpha, \beta\}$ & $\max \{\alpha, \beta\}$
    & $1 - \alpha$ & $\max \{1 - \alpha, \beta\}$ \\
    \hline
    G\"{o}del & $\min \{\alpha, \beta\}$ & $\max \{\alpha, \beta\}$ &
    $\left\{
    \begin{array}{ll}
    1, & \alpha = 0 \\
    0, & \alpha > 0 \\
    \end{array}
    \right.$ &
    $\left\{
    \begin{array}{ll}
    1 & \alpha \leq \beta \\
    \beta, & \alpha > \beta \\
    \end{array}
    \right.$\\
    \hline
    {\L}ukasiewicz & $\max \{\alpha + \beta - 1, 0\}$ & $\min \{\alpha + \beta, 1\}$ & $1 - \alpha$ &
    $\min \{1 - \alpha + \beta, 1\}$ \\
%
    \hline
    Product & $\alpha \cdot \beta$ & $\alpha + \beta - \alpha \cdot \beta$ &
    $\left\{
    \begin{array}{ll}
    1, & \alpha = 0 \\
    0, & \alpha > 0 \\
    \end{array}
    \right.$ &
    $\left\{
    \begin{array}{ll}
    1 & \alpha \leq \beta \\
    \beta / \alpha, & \alpha > \beta \\
    \end{array}
    \right.$\\
    \hline
\end{tabular}
} \hss }
\end{table}

A fuzzy set $C$ is \emph{included} in another fuzzy set $D$ iff
$\forall x \in X, \mu_{C}(x) \leq \mu_{D}(x)$. According to this
definition, which is usually called Zadeh's set inclusion, fuzzy set
inclusion is a yes-no question. In order to overcome this, other
definitions have been proposed. For example, the degree of inclusion
of $C$ in $D$ can be computed using some implication function as
$\inf_{x \in X} \mu_{C}(x) \Rightarrow \mu_{D}(x)$. Note that these
two approaches are equivalent under Rescher implication, defined as
$\alpha \Rightarrow \beta = 1$ iff $\alpha \leq \beta$, or $\alpha
\Rightarrow \beta = 0$ otherwise.

A (binary) \emph{fuzzy relation} $R$ over two countable classical
sets $X$ and $Y$ is a function $R \colon X\times Y \to \unit$. The
\emph{inverse} of $R$ is the function $R^{-1}\colon Y \times X \to
\unit$ with membership function $R^{-1}(y,x) = R(x,y)$, for every $x
\in X$ and $y \in Y$. The \emph{composition} of two fuzzy relations
$R_{1}\colon X\times Y \to \unit$ and $R_{2}\colon Y\times Z \to
\unit$ is defined as $(R_{1} \circ R_{2})(x,z) = \sup_{y\in Y} R_{1}
(x,y) \otimes R_{2}(y,z)$. A~fuzzy relation $R$ is \emph{transitive}
iff  $R(x,z) \ges (R\circ R)(x,z)$.

A fuzzy interpretation $\I$  \emph{satisfies} a fuzzy statement
$\phi \ges l$ (resp., $\phi \les u$) or  $\I$ is a \emph{model} of
$\phi \ges l$ (resp., $\phi \les u$), denoted $\I\,{\models}\,\phi
\ges l$ (resp., $\I\,{\models}\,\phi \les u$), iff $\I(\phi) \ges l$
(resp., $\I(\phi) \les u$). The notions of satisfiability and
logical consequence are defined in the standard way. We say
that~$\phi\ges l$ is a {\em tight logical consequence} of a set of
fuzzy statements $\KB$ iff $l$ is the infimum of~$\I(\phi)$ subject
to all models~$\I$ of~$\KB$. Notice that the latter is equivalent to
$l\eqs \sup \,\{ r \mids \KB \,{\models}\, \phi\ges r \}$. For
reasoning algorithms for fuzzy propositional and First-Order Logics
see~\cite{Hajek98}.


\section{The Fuzzy DL $\mathcal{SROIQ}\mathbf{(D)}$}
~\label{sec:fuzzySROIQD}

\nd In this section we describe the fuzzy DL
$\mathcal{SROIQ}\mathbf{(D)}$, a subset of the language presented
in~\cite{BobilloFuzzIEEE2010a}, which was inspired by the logics
presented in~\cite{BobilloIJAR,BobilloFuzzIEEE2008,StracciaSHOIND}.
Here, concepts denote fuzzy sets of individuals and roles denote
fuzzy binary relations. Axioms are also extended to the fuzzy case
and some of them hold to a degree.

\subsection{Syntax}
\label{sec:syntax}

\paragraph{Notation}
To begin with, we will introduce some notation that will be used in
the rest of the paper:

{\footnotesize
\begin{itemize}
\item $C,D$ are (possibly complex) fuzzy concepts,
\item $A$ is an atomic fuzzy concept,
\item $R$ is a (possibly complex) abstract fuzzy role,
\item $R_A$ is an atomic fuzzy role,
\item $S$ is a simple fuzzy role~\footnote{Simple
roles are needed to guarantee the decidability of the logic.
Intuitively, simple roles cannot take part in cyclic role inclusion
axioms (see~\cite{BobilloIJUFKS} for a formal definition).},
\item $T$ is a concrete fuzzy role,
\item $a,b$ are abstract individuals, $v$ is a concrete individual,
\item $\mathbf{d}$ is a fuzzy concrete predicate,
\item $n, m$ are natural numbers with $n \ges 0, m > 0$,
\item $mod$ is a fuzzy modifier,
\item $\rhd \in \{ \geq, > \}, \lhd \in \{ \leq, < \}, \bowtie\;\in \{\geq, >, \leq, < \}$,
\item $\alpha \in [0,1]$.
\end{itemize}
}

\nd Next, we will introduce two important elements of our logic:
fuzzy modifiers and fuzzy concrete domains.

\begin{figure}[htbp]
\begin{center}
\begin{tabular}{ccccc}
\includegraphics[scale=0.2]{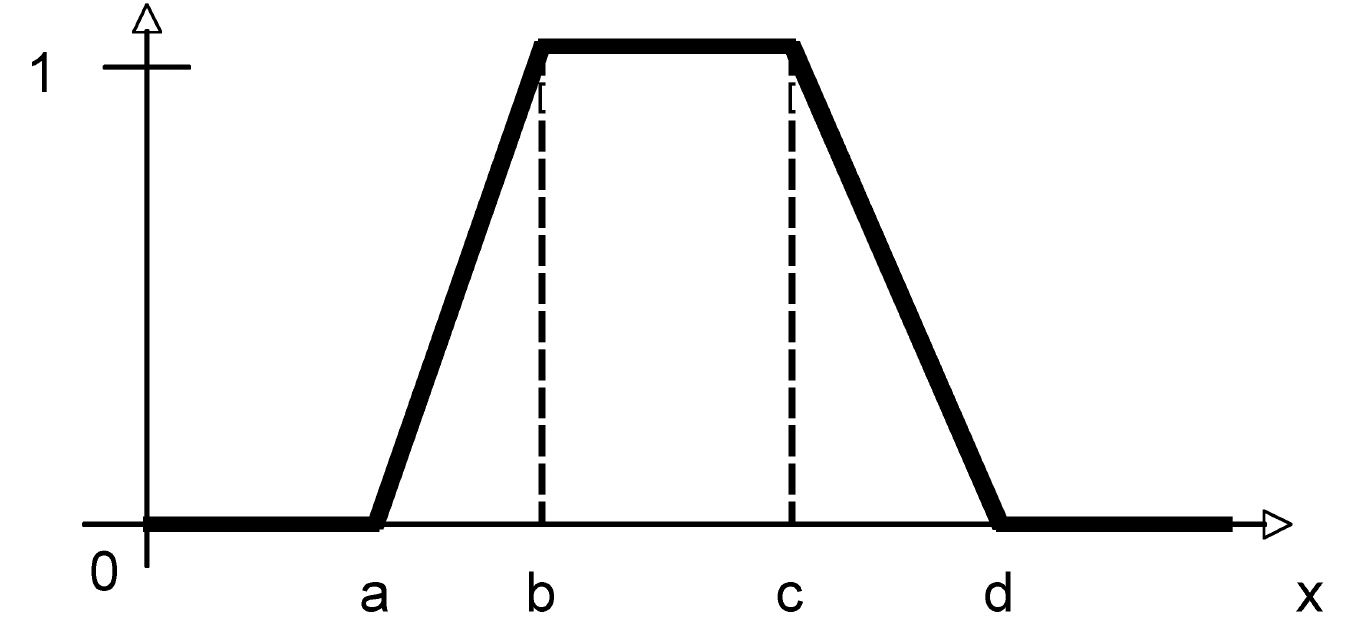}&
\includegraphics[scale=0.2]{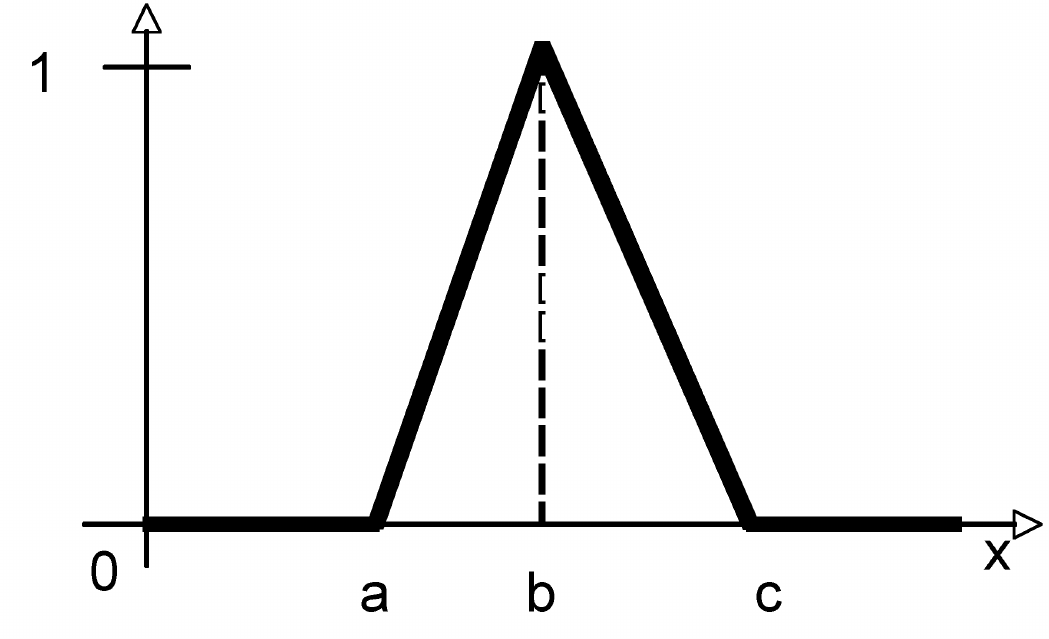} &
\includegraphics[scale=0.2]{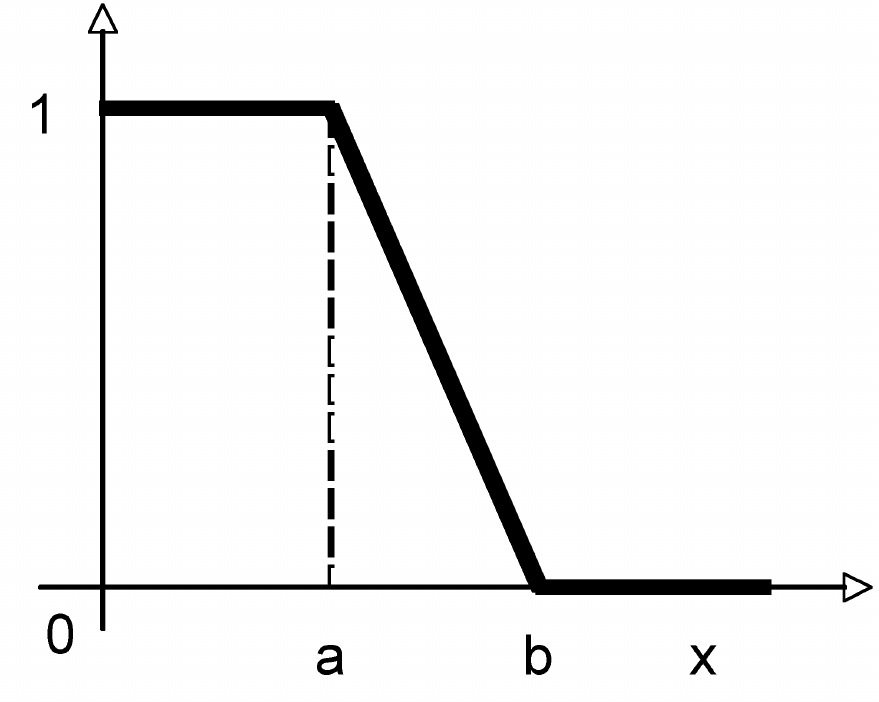} &
\includegraphics[scale=0.2]{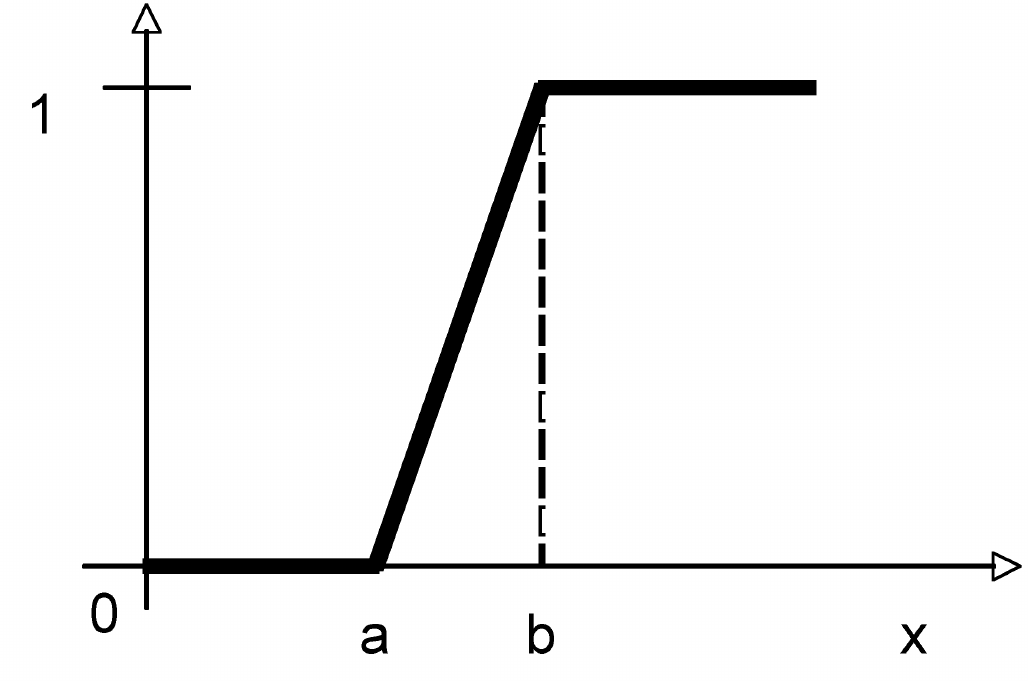} &
\includegraphics[scale=0.3]{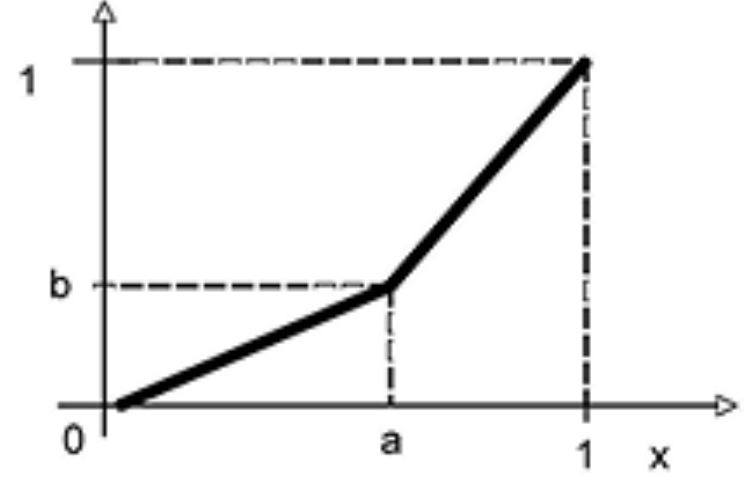} \\
(a) & (b) & (c) & (d) & (e)  \\
\end{tabular}
\caption{(a) Trapezoidal function; (b) Triangular function; (c)
$L$-function; (d) $R$-function; (e) Linear
function.}\label{fig:membFunctions}
\end{center}
\end{figure}

\paragraph{Fuzzy modifiers}
A \emph{fuzzy modifier} $mod$ is a function $f_{mod}\colon\unit \to
\unit$ which applies to a fuzzy set to change its membership
function. We will allow modifiers defined in terms of \emph{linear}
hedges (Figure~\ref{fig:membFunctions} (e)) and \emph{triangular}
functions (Figure~\ref{fig:membFunctions} (b))~\cite{StracciaALCD1}. Formally: \\

{\small
\begin{tabular}{rrcl}
$mod \to$ & $\mathtt{linear}(c)$       & $|$ & (M1) \\
          & $\mathtt{triangular}(a,b,c)$ &   & (M2) \\
\end{tabular}
} \\

\noindent where in linear modifiers we assume that $a = c/(c+1), b =
1/(c+1)$.

\begin{example}
\label{ex:fuzzyModifier} Modifier $\textsf{very}$ can be defined as
$\mathtt{linear}(0.8)$.
\end{example}

\paragraph{Fuzzy concrete domains}
\label{sec:fuzzyConcreteDomain}

A \emph{fuzzy concrete domain}~\cite{StracciaALCD1} (also called a
fuzzy \emph{datatype}) $\mathbf{D}$ is a pair $\langle
\Delta_{\mathbf{D}}, \Phi_{\mathbf{D}} \rangle$, where
$\Delta_{\mathbf{D}}$ is a concrete interpretation domain, and
$\Phi_{\mathbf{D}}$ is a set of fuzzy concrete predicates
$\mathbf{d}$ with an arity $n$ and an interpretation
$\mathbf{d}^{\mathcal{I}} : \Delta^n_{\mathbf{D}} \to \unit$, which
is an $n$-ary fuzzy relation over $\Delta_{\mathbf{D}}$.

As fuzzy concrete predicates we allow the following functions
defined over an interval $[k_1,k_2] \subseteq \mathbb{Q}$:
\emph{trapezoidal} membership function
(Figure~\ref{fig:membFunctions} (a)), the \emph{triangular}
(Figure~\ref{fig:membFunctions} (b)), the \emph{left-shoulder}
function (Figure~\ref{fig:membFunctions} (c)) and the
\emph{right-shoulder} function (Figure~\ref{fig:membFunctions}
(d))~\cite{StracciaALCD1}.

Furthermore, we will also allow \emph{fuzzy modified datatypes},
obtained after the application of a fuzzy modifier $mod$ to a fuzzy
concrete domain interpretation.

Formally: \\

{\small
\begin{tabular}{rrcl}
$\mathbf{d} \to$ &  $\mathtt{left}(k_1,k_2,a,b)$ & $|$ & (D1) \\
    & $\mathtt{right}(k_1,k_2,a,b)$ & $|$ & (D2) \\
    & $\mathtt{triangular}(k_1,k_2,a,b,c)$ & $|$ & (D3) \\
    & $\mathtt{trapezoidal}(k_1,k_2,a,b,c,d)$ & $|$   & (D4) \\
    & $mod(\mathbf{d})$ &   & (D5) \\
\end{tabular}
}

\noindent Note that in fuzzy modified datatypes $k_1 = 0, k_2 = 1$.
Furthermore, we allow nesting of modifiers, as for example
$mod(mod(d))$.

\begin{example}
\label{ex:fuzzyDatatype} We may define the fuzzy datatype
$\textsf{YoungAge} \colon [0, 200] \to \unit$, denoting the degree
of a person being young, as $\textsf{YoungAge}(x) = \mathtt{left}(0,
200, 10,30)$.
\end{example}

\paragraph{Symbols}
Fuzzy $\mathcal{SROIQ}\mathbf{(D)}$ assumes three alphabets of
symbols, for (abstract and concrete) \emph{fuzzy concepts},
\emph{fuzzy roles} and \emph{individuals}. The syntax of fuzzy
concepts and roles is shown in Table~\ref{tab:semantics}.

Concept constructors (C1)--(C16) correspond to the concept
constructors of crisp $\mathcal{SROIQ}\mathbf{(D)}$. The only
difference here are modified concepts (C17), weighted concepts
(C18), and weighted sum concepts (C19). In (C19), se assume that
$\sum_{i=1}^{k} \alpha_i \leq 1$.

\begin{example}
\label{ex:concepts} Concept $\textsf{Human} \sqcap \exists
\textsf{hasAge}.\textsf{YoungAge}$ denotes the fuzzy set of young
humans. $\textsf{very}(\textsf{Human} \sqcap \exists
\textsf{hasAge}.\textsf{YoungAge})$ denotes \emph{very} young
humans.
\end{example}

\nd Role constructors (R1)--(R3) correspond to the role constructors
of crisp $\mathcal{SROIQ}\mathbf{(D)}$. (R4) corresponds to modified
roles.

\paragraph{Fuzzy Knowledge Base}
A \emph{Fuzzy Knowledge Base} (KB) contains a finite number of
axioms. The axioms that are allowed in our logic are shown in
Table~\ref{tab:semantics}. They can be grouped into a fuzzy ABox
with axioms (A1)--(A7), a fuzzy TBox with axioms (A8)--(A11), and a
fuzzy RBox with axioms (A12)--(A25). All the axioms have a
equivalent in crisp  $\mathcal{SROIQ}\mathbf{(D)}$.

\begin{example}
\label{ex:axioms} The fuzzy concept assertion $\langle \textsf{paul}
\colon \textsf{Tall} \geq 0.5 \rangle$ states that Paul is tall with
at least degree $0.5$. The fuzzy RIA $\langle \textsf{isFriendOf}$
$\textsf{isFriendOf} \sqsubseteq \textsf{isFriendOf} \geq 0.75
\rangle$ states that the friends of my friends can also be
considered as my friends with at least degree $0.75$.
\end{example}


\subsection{Semantics}
\label{sec:semantics}

\paragraph{Fuzzy interpretation}
A fuzzy interpretation $\mathcal{I}$ with respect to a fuzzy
concrete domain $\mathbf{D}$ is a pair $(\Delta^{\mathcal{I}}, \cdot
^{\mathcal{I}})$ consisting of a non empty set
$\Delta^{\mathcal{I}}$ (the interpretation domain) disjoint with
$\Delta_{\mathbf{D}}$ and a fuzzy interpretation function $\cdot
^{\mathcal{I}}$ mapping:

\begin{itemize}

\item A fuzzy \emph{abstract individual} $a$ onto an element $a^{\mathcal{I}} \subseteq \Delta^{\mathcal{I}}$.

\item A fuzzy \emph{concrete individual} $v$ onto an element $v_{\mathbf{D}} \subseteq \Delta_{\mathbf{D}}$.

\item A fuzzy \emph{concept} $C$ onto a function $C^{\mathcal{I}}:
\Delta^{\mathcal{I}} \to \unit$.

\item A fuzzy \emph{abstract role} $R$ onto a function
$R^{\mathcal{I}}: \Delta^{\mathcal{I}} \times \Delta^{\mathcal{I}}
\to \unit$.

\item A fuzzy \emph{concrete role} $T$ onto a function
$T^{\mathcal{I}}: \Delta^{\mathcal{I}} \times \Delta_{\mathbf{D}}
\to \unit$.


\item An $n$-ary fuzzy \emph{concrete domain}
$\mathbf{d}$ onto a function $\mathbf{d}^{\mathcal{I}} :
\Delta^{n}_{\mathbf{D}} \to \unit$.

\item A fuzzy \emph{modifier} $mod$ onto a function $f_{mod} : \unit \to \unit]$.

\end{itemize}

\nd $C^{\mathcal{I}}$ (resp. $R^{\mathcal{I}}$) denotes the membership
function of the fuzzy concept $C$ (resp. fuzzy role $R$) w.r.t.
$\mathcal{I}$. $C^{\mathcal{I}}(a)$ (resp. $R^{\mathcal{I}}(a,b)$)
gives us to what extent the individual $a$ can be considered as an
element of the fuzzy concept $C$ (resp. to what extent $(a,b)$ can
be considered as an element of the fuzzy role $R$) under the fuzzy
interpretation $\mathcal{I}$.

The fuzzy interpretation function is defined for fuzzy concepts,
roles, concrete domains and axioms as shown in
Table~\ref{tab:semantics}. We say that a fuzzy interpretation $\I$
satisfies a fuzzy KB $\mathcal{K}$ iff $\I$ satisfies each element
in $\mathcal{K}$.

\begin{table}[htbp]
\caption{Syntax and semantics of the fuzzy DL
$\mathcal{SROIQ}\mathbf{(D)}$.} \label{tab:semantics}
\begin{center}{\scriptsize
\begin{tabular}{lll}
  \hline
  \textbf{Concept} & \textbf{Syntax ($C$)} & \textbf{Semantics of $C^{\mathcal{I}}(x)$} \\
  \hline
  (C1) & $A$ &  $A^{\mathcal{I}}(x)$ \\
  (C2) & $\top$ &  $1$ \\
  (C3) & $\bot$ &   $0$ \\
  (C4) & $C \sqcap D$ &  $C^{\mathcal{I}}(x) \otimes D^{\mathcal{I}}(x)$ \\
  (C5) & $C \sqcup D$ &  $C^{\mathcal{I}}(x) \oplus D^{\mathcal{I}}(x)$ \\
  (C6) & $\neg C$ &  $\ominus C^{\mathcal{I}}(x)$ \\
  (C7) & $\forall R.C$ &  $\inf_{y \in \Delta^{\mathcal{I}}} \{ R^{\mathcal{I}}(x,y) \Rightarrow C^{\mathcal{I}}(y) \}$ \\
  (C8) & $\exists R.C$ &  $\sup_{y \in \Delta^{\mathcal{I}}} \{ R^{\mathcal{I}}(x,y) \otimes C^{\mathcal{I}}(y) \}$ \\
  (C9) & $\forall T.\mathbf{d}$ &  $\inf_{v \in \Delta_{\mathbf{D}}} \{ T^{\mathcal{I}}(x,v) \Rightarrow \mathbf{d}^{\mathcal{I}}(v) \}$ \\
  (C10) & $\exists T.\mathbf{d}$ &  $\sup_{v \in \Delta_{\mathbf{D}}} \{ T^{\mathcal{I}}(x,v) \otimes \mathbf{d}^{\mathcal{I}}(v) \}$ \\
  (C11) & $\{\alpha / a \}$ &  $\alpha$ if $x = o_{i}^{\mathcal{I}}$, $0$ otherwise \\
  (C12) & $\geq m\;S.C$ &  $\sup_{y_{1}, \dots, y_{m} \in \Delta^{\mathcal{I}}} { (\min^{m}_{i=1} \{S^{\mathcal{I}}(x,y_i) \otimes C^{\mathcal{I}}(y_i) \}) \bigotimes ((\otimes)_{1 \leq j < k \leq m} \{ y_{j} \neq y_{k}\})} $ \\
  (C13) & $\leq n\;S.C$ &  $\inf_{y_{1}, \dots, y_{n+1} \in \Delta^{\mathcal{I}}} { (\min^{n+1}_{i=1} \{S^{\mathcal{I}}(x,y_i) \otimes C^{\mathcal{I}}(y_i) \}) \Rightarrow ((\oplus)_{1 \leq j < k \leq n+1} \{ y_{j} = y_{k}\})} $ \\
  (C14) & $\geq m\;T.\mathbf{d}$ &  $\sup_{v_{1}, \dots, v_{m} \in \Delta_{\mathbf{D}}} { (\min^{m}_{i=1} \{T^{\mathcal{I}}(x,v_{i}) \otimes \mathbf{d}^{\mathcal{I}}(v_{i}) \}) \bigotimes ((\otimes)_{j < k} \{ v_{j} \neq v_{k}\})} $ \\
  (C15) & $\leq n\;T.\mathbf{d})$ &  $\inf_{v_{1}, \dots, v_{n+1} \in \Delta_{\mathbf{D}}} { (\min^{n+1}_{i=1} \{T^{\mathcal{I}}(x,v_{i}) \otimes \mathbf{d}^{\mathcal{I}}(v_{i}) \}) \Rightarrow ((\oplus)_{j < k} \{ v_{j} = v_{k}\})} $ \\
  (C16) & $\exists S.\texttt{Self}$ &  $S^{\mathcal{I}}(x,x)$ \\
  (C17) & $mod(C)$ &  $f_{mod}(C^{\mathcal{I}}(x))$ \\
  (C18) & $\alpha \cdot C$ &  $\alpha \cdot C^{\mathcal{I}}(x)$ \\
  (C19) & $(\alpha_1 \cdot C_1) + \dots + (\alpha_k \cdot C_k)$ &  $\sum^{k}_{i=1} \alpha_i \cdot C_i^{\mathcal{I}}(x)$ \\
  \hline
  \textbf{Role} & \textbf{Syntax ($R$)} & \textbf{Semantics of $R^{\mathcal{I}}(x,y)$} \\
  \hline
  (R1) & $R_A$ &  $R_A^{\mathcal{I}}(x,y)$ \\
  (R2) & $R^-$ &  $R^{\mathcal{I}}(y,x)$ \\
  (R3) & $U$ &  $1$ \\
  (R4) & $mod(R)$ &  $f_{mod}(R^{\mathcal{I}}(x,y))$ \\
  (R5) & $T$ &  $T^{\mathcal{I}}(x,y)$ \\
  \hline
  \textbf{Datatype} & \textbf{Syntax ($\mathbf{d}$)} & \textbf{Semantics of $\mathbf{d}^{\mathcal{I}}$} \\
  \hline
  (D1--D4) & See Section~\ref{sec:fuzzyConcreteDomain} & $\mathbf{d}_{\mathbf{D}}$ \\
  (D5)     & $mod(\mathbf{d})$ &  $f_{mod}(\mathbf{d}^{\mathcal{I}})$ \\
  \hline
  \textbf{Axiom} & \textbf{Syntax ($\tau$)} & \textbf{Semantics} ($\I$ satisfies $\tau$ if \dots) \\
  \hline
  (A1) & $\langle a\!:\!C \bowtie \alpha \rangle$ &  $C^{\mathcal{I}}(a^{\mathcal{I}}) \bowtie \alpha$ \\
  (A2) & $\langle (a,b)\!:\!R \bowtie \alpha \rangle$ &  $R^{\mathcal{I}}(a^{\mathcal{I}},b^{\mathcal{I}}) \bowtie \alpha$ \\
  (A3) & $\langle (a,b)\!:\!\neg R \bowtie \alpha \rangle$ &  $\ominus R^{\mathcal{I}}(a^{\mathcal{I}},b^{\mathcal{I}}) \bowtie \alpha$ \\
  (A4) & $\langle (a,v)\!:\!T \bowtie \alpha \rangle$ &  $T^{\mathcal{I}}(a^{\mathcal{I}},v_{\mathbf{D}}) \bowtie \alpha$ \\
  (A5) & $\langle (a,v)\!:\!\neg T \bowtie \alpha \rangle$ &  $\ominus T^{\mathcal{I}}(a^{\mathcal{I}},v_{\mathbf{D}}) \bowtie \alpha$ \\
  (A6) & $\langle a \not= b \rangle$ &  $a^{\mathcal{I}} \neq b^{\mathcal{I}}$ \\
  (A7) & $\langle a = b \rangle$ &  $a^{\mathcal{I}} = b^{\mathcal{I}}$ \\
  (A8) & $\langle C \sqsubseteq D \rhd \alpha \rangle$ &  $\inf_{x \in \Delta^{\mathcal{I}}} \{ C^{\mathcal{I}}(x) \Rightarrow D^{\mathcal{I}}(x) \} \rhd \alpha$ \\
  (A9) & $C_1 \equiv \dots C_m$ &  $\forall_{x \in \Delta^{\mathcal{I}}} C_1^{\mathcal{I}}(x) = \dots = C_m^{\mathcal{I}}(x)$ \\
  (A10) & $\texttt{dis}(C_{1}, \dots, C_{m})$ &  $\forall x,y \in \Delta^{\mathcal{I}}, \min \{ C_{1}^{\mathcal{I}}(x,y), \dots, C_{m}^{\mathcal{I}}(x,y) \} = 0$ \\
  (A11) & $\texttt{disUnion}(C_{1}, \dots, C_{m})$ &  $\texttt{dis}(C_{2}, \dots, C_{m}), C_1 \equiv C_2 \sqcup \dots \sqcup C_m $ \\
  (A12) & $\langle R_{1} \dots R_{m} \sqsubseteq R \rhd \alpha \rangle$ &  $\inf_{x_{1}, x_{n+1} \in \Delta^{\mathcal{I}}} \{ \sup_{ x_2 \dots x_{m} \in \Delta^{\mathcal{I}}} \{( R^{\mathcal{I}}_{1}(x_1,x_2) \otimes \dots \otimes R^{\mathcal{I}}_{n}(x_m,x_{m+1}) ) \Rightarrow $ \\
       & & $  R^{\mathcal{I}}(x_1,x_{m+1}) \} \} \rhd \alpha$ \\
  (A13) & $\langle T_1 \sqsubseteq T_2 \rhd \alpha \rangle$ &  $\inf_{x \in \Delta^{\mathcal{I}}, v \in \Delta_{\mathbf{D}}} \{ T_1^{\mathcal{I}}(x,v) \Rightarrow T_2^{\mathcal{I}}(x,v) \} \rhd \alpha$ \\
  (A14) & $R_1 \equiv \dots R_m$ &  $\forall_{x,y \in \Delta^{\mathcal{I}}} R_1^{\mathcal{I}}(x,y) = \dots = R_m^{\mathcal{I}}(x,y)$ \\
  (A15) & $T_1 \equiv \dots T_m$ &  $\forall_{x \in \Delta^{\mathcal{I}}, v \in \Delta_{\mathbf{D}}} R_1^{\mathcal{I}}(x,v) = \dots = R_m^{\mathcal{I}}(x,v)$ \\
  (A16) & $\texttt{domain}(R,C)$ &  $\langle \exists R.\top \sqsubseteq C \geq 1 \rangle$ \\
  (A17) & $\texttt{range}(R,C)$ &  $\langle \top \sqsubseteq \forall R.C \geq 1 \rangle$ \\
  (A18) & $\texttt{func}(R)$ &  $\langle \top \sqsubseteq (\leq 1 \; R.\top) \geq 1 \rangle$ \\
  (A19) & $\texttt{trans}(R)$ &  $\forall x,y,z \in \Delta^{\mathcal{I}}, R^{\mathcal{I}}(x,z) \otimes R^{\mathcal{I}}(z,y) \leq R^{\mathcal{I}}(x,y)$ \\
  (A20) & $\texttt{dis}(S_{1}, \dots, S_{m})$ &  $\forall x,y \in \Delta^{\mathcal{I}}, \min \{ S_{1}^{\mathcal{I}}(x,y), \dots, S_{m}^{\mathcal{I}}(x,y) \} = 0$ \\
  (A21) & $\texttt{dis}(T_{1}, \dots, T_{m})$ &  $\forall x \in \Delta^{\mathcal{I}}, v \in \Delta_{\mathbf{D}}, \min \{ T_{1}^{\mathcal{I}}(x,v), \dots, T_{m}^{\mathcal{I}}(x,v) \} = 0$ \\
  (A22) & $\texttt{ref}(R)$ &  $\forall x \in \Delta^{\mathcal{I}}, R^{\mathcal{I}}(x,x) = 1$ \\
  (A23) & $\texttt{irr}(S)$ &  $\forall x \in \Delta^{\mathcal{I}}, S^{\mathcal{I}}(x,x) = 0$ \\
  (A24) & $\texttt{sym}(R)$ &  $\forall x,y \in \Delta^{\mathcal{I}}, R^{\mathcal{I}}(x,y) = R^{\mathcal{I}}(y,x)$ \\
  (A25) & $\texttt{asy}(S)$ &  $\forall x,y \in \Delta^{\mathcal{I}} $, if $S^{\mathcal{I}}(x,y) > 0$ then $S^{\mathcal{I}}(y,x) = 0$ \\
\end{tabular}
}
\end{center}
\end{table}

Note that we have included some syntactic sugar axioms: concept
equivalences (A9), disjoint concept axioms (A10), disjoint union
concepts (A11), domain role axioms (A16), range role restrictions
(A17), and functional role axioms (A18). In fact, while in the
classical case the meaning of these axioms is very clear, in the
fuzzy case this is not always the case. As discussed
in~\cite{StoilosIJAR}, there could be alternative definitions for
disjoint concepts, and range role axioms. Consequently, it was
convenient to write the formal definition of these axioms.

\subsection{Reasoning tasks}

\nd There are several reasoning tasks in fuzzy
$\mathcal{SROIQ}\mathbf{(D)}$~\cite{Straccia01,StracciaSHOIND}.

\begin{itemize}

    \item \emph{Fuzzy KB satisfiability}. A fuzzy interpretation $\I$ \emph{satisfies}
    (is a model of) a fuzzy KB $\KB$  iff it satisfies each axiom in $\KB$.

    \item \emph{Concept satisfiability}. $C$ is $\alpha$-satisfiable w.r.t. a fuzzy KB $\KB$ iff
    there exists a model $\I$ of $\KB$ such that $C^I(x) \geq \alpha$ for some $x \in \Delta^\I$.

    \item \emph{Entailment}: A fuzzy concept (or role) assertion $\tau$ is entailed by a fuzzy KB $\KB$
    iff every model of $\KB$ satisfies $\tau$.

    \item \emph{Concept subsumption}: $D$ subsumes $C$ (denoted $C \sqsubseteq D$) w.r.t. a fuzzy KB $\KB$
    iff every model $\I$ of $\KB$ satisfies $\forall x \in \Delta^\I, C^\I(x) \leq D^\I(x)$.

    \item \emph{Best degree bound} (BDB). The BDB of a concept or role assertion $\tau$
    is defined as the $\sup \{\alpha : \KB \models \langle \tau \geq \alpha \rangle \}$.

    \item \emph{Maximal concept satisfiability degree}. The maximal satisfiability degree of a fuzzy concept $C$ w.r.t. a fuzzy KB $\KB$
    is defined as the $\sup \{ \alpha | C$ is $\alpha$-satisfiable $\}$.

\end{itemize}

\nd However, these reasoning tasks are part of the query language and
not of the representation language. Thus, we shall not represent
them in a fuzzy ontology.


\section{Representation of Fuzzy Ontologies in OWL 2}
\label{sec:owl2}

\nd In this section we will explain a methodology to represent fuzzy
$\mathcal{SROIQ}\mathbf{(D)}$ ontologies using OWL 2. We anticipate
that the methodology has some differences with a previous version in
the paper~\cite{BobilloFuzzIEEE2010a}, as explained in Section
~\ref{sec:discussion}.

The idea of our representation is to use an OWL 2 ontology,
extending their elements with annotation properties representing the
features of the fuzzy ontology that OWL 2 cannot directly encode.

For the sake of clarity, we will use OWL 2 abstract
syntax~\cite{OWL2syntax} for OWL 2, and an XML syntax to write the
value of annotation properties\footnote{Of course, the final result
depends on the syntax (for instance, in OWL 2 XML syntax the
characters $\geq$ and $\leq$ of the annotations are escaped), but
OWL 2 ontology editors make these issues transparent to the user.}.

Let us begin with an illustrating example.

\begin{example}
\label{ex:annotation} Consider the fuzzy concept assertion of
Example~\ref{ex:axioms}, $\langle \textsf{paul} \colon \textsf{Tall}
\geq 0.5 \rangle$. To represent it in OWL 2, we consider the crisp
assertion $\textsf{paul} \colon \textsf{Tall}$ as represented in OWL
2, \verb"ClassAssertion(paul Tall)" and then we add an annotation
property including the information $\geq 0.5$ to it.
\end{example}

\nd It is worth to note that OWL 2 only provides for annotations on
ontologies, axioms, and entities~\cite{OWL2syntax}. This is not the
case of OWL DL, which only provides for annotations on ontologies
and entities.

\subsection{Syntactic Requirements of Fuzzy Ontologies}

\nd To begin with, we will summarize the syntactic differences between
the fuzzy and non-fuzzy ontologies. There are $6$ cases depending on
the annotated element.

\begin{description}

\item[Case 1.]  Fuzzy modifiers do not have an equivalence in the non-fuzzy case: (M1), (M2).

\item[Case 2.]  Fuzzy datatypes do not have an equivalence in the non-fuzzy case: (D1)--(D5).

\item[Case 3.]  Some fuzzy concepts have syntactic differences with the non-fuzzy case (C11)
or do not have an equivalence (C17)--(C19).

\item[Case 4.]  Some fuzzy roles do not have an equivalence in the non-fuzzy case: (R4).

\item[Case 5.]  Some axioms require an inequality sign and a degree of truth:
(A1)--(A5), (A8), (A12)--(A13).

\item[Case 6.]  Ontologies can be annotated with a fuzzy logic.

\end{description}

\subsection{Annotations}

\nd Instead of using any of the defaults annotation properties from OWL
2, we will use an annotation property \verb"fuzzyLabel".
Furthermore, for every element of the ontology there can be at-most
one annotation of this type.

Every annotation will be delimited by a start tag \verb"<FuzzyOwl2>"
and an end tag \verb"</FuzzyOwl2>", with an attribute
\verb"fuzzyType" specifying the fuzzy element being tagged. In the
following, we will address the different cases in detail.

\subsection{Fuzzy modifiers}

\nd According to Section~\ref{sec:syntax}, the fuzzy modifiers that
we want to represent have parameters $a,b,c$. Consequently, they can
be represented as in the previous case, with the particularities
that the type of datatype should be double (\verb"xsd:double") and
that there is no need to use \verb"xsd:minInclusive" and
\verb"xsd:maxInclusive" (they are assumed to be $0, 1$).

The value of \verb"fuzzyType" will be \verb"modifier", and there
will be a tag \verb"Modifier" with an attribute \verb"type"
(possible values \verb"linear", and \verb"triangular"), and
attributes \verb"a", \verb"b", \verb"c", depending on the type of
the modifier.

\paragraph{Domain of the annotation}
An OWL 2 datatype declaration of the type base double
\verb"xsd:double".

\paragraph{Syntax for the annotation} \textsc{ }

\lstset{basicstyle=\ttfamily \scriptsize}
\begin{lstlisting}[frame=none]{}
<fuzzyOwl2 fuzzyType="modifier">
  <MODIFIER>
</fuzzyOwl2>

<MODIFIER> :=
  <Modifier type="linear" c="<DOUBLE>" /> |
  <Modifier type="triangular" a="<DOUBLE>" b="<DOUBLE>" c="<DOUBLE>" />
\end{lstlisting}

\paragraph{Semantical restrictions}
The parsers should check that the following constraints:

\begin{itemize}
    \item $a, b, c \in \unit$
    \item $b = 0$ iff $a = 1$
    \item $b = 1$ iff $c = 1$
\end{itemize}

\begin{example}
\label{ex:modifier}

Let us define the fuzzy modifier $\textsf{Very} =
\mathtt{linear}(0.8)$. We create a datatype \textsf{Very}.

\lstset{basicstyle=\ttfamily \scriptsize}
\begin{lstlisting}[frame=none]{}
DatatypeDefinition ( Very DatatypeRestriction (
  xsd:double
  xsd:minInclusive "0"^^xsd:double
  xsd:maxInclusive "1"^^xsd:double
) )
\end{lstlisting}

\nd Then, we add the following annotation property to it:

\lstset{basicstyle=\ttfamily \scriptsize}
\begin{lstlisting}[frame=none]{}
<fuzzyOwl2 fuzzyType="modifier">
  <Modifier type="linear" c="0.8" />
</fuzzyOwl2>
\end{lstlisting}

\end{example}

\subsection{Fuzzy datatypes}

\nd Firstly, we will consider fuzzy datatypes (D1)--(D4), and then
we will consider the case (D5).

\subsubsection{Fuzzy atomic datatypes}

\nd According to Section~\ref{sec:syntax}, these fuzzy datatypes
have parameters $k_1,k_2,a,b,c,d$. The first four parameters are
common to all of them, $c$ only appears in (D4), (D5); and $d$ only
appears in (D5).

\paragraph{Domain of the annotation}
An OWL 2 datatype declaration of the type base of the fuzzy datatype
(integer \verb"xsd:integer" or double \verb"xsd:double"), such that:

\lstset{basicstyle=\ttfamily \scriptsize}
\begin{lstlisting}[frame=none]{}
xsd:minInclusive="<DOUBLE>"
xsd:maxInclusive="<DOUBLE>"
\end{lstlisting}

\nd \verb"<DOUBLE>" denotes a rational number.
\verb"xsd:minInclusive" should take the value $k_1$, whereas
\verb"xsd:maxInclusive" should take the value $k_2$. These
parameters are optional and, if omitted, then the minimum and
maximum of the attributes ($a, b, c, d$) is assumed, respectively.

\paragraph{Syntax for the annotation} \textsc{ }

\lstset{basicstyle=\ttfamily \scriptsize}
\begin{lstlisting}[frame=none]{}
<fuzzyOwl2 fuzzyType="datatype">
  <DATATYPE>
</fuzzyOwl2>

<DATATYPE> :=
  <Datatype type="leftshoulder" a="<DOUBLE>" b="<DOUBLE>" /> |
  <Datatype type="rightshoulder" a="<DOUBLE>" b="<DOUBLE>" /> |
  <Datatype type="triangular" a="<DOUBLE>" b="<DOUBLE>" c="<DOUBLE>" /> |
  <Datatype type="trapezoidal" a="<DOUBLE>" b="<DOUBLE>" c="<DOUBLE>" d="<DOUBLE>" />
\end{lstlisting}

\paragraph{Semantical restrictions}
The parsers should check the following restrictions:

\begin{itemize}
    \item $k_1 \leq a \leq b \leq c \leq d \leq k_2$ is verified.
\end{itemize}

\begin{example}
\label{ex:datatype}
Let us represent the fuzzy datatype $\textsf{YoungAge} =
\mathtt{left}(0, 200, 10,30)$ denoting the age of a young person.
This fuzzy datatype is represented using a datatype definition of
base type \verb"xsd:integer" with range in $[0, 200]$:

\lstset{basicstyle=\ttfamily \scriptsize}
\begin{lstlisting}[frame=none]{}
DatatypeDefinition ( YoungAge DatatypeRestriction (
  xsd:integer
  xsd:minInclusive "0"^^xsd:integer
  xsd:maxInclusive "200"^^xsd:integer
) )
\end{lstlisting}

\nd Then we add the following annotation property to it:

\lstset{basicstyle=\ttfamily \scriptsize}
\begin{lstlisting}[frame=none]{}
<fuzzyOwl2 fuzzyType="datatype">
  <Datatype type="leftshoulder" a="10" b="30" />
</fuzzyOwl2>
\end{lstlisting}

\end{example}

\subsubsection{Fuzzy modified datatypes}

\nd In this case, the parameters are two: the modifier, and the
fuzzy datatype that is being modified.

\paragraph{Domain of the annotation}
An OWL 2 datatype declaration of any type base.

\paragraph{Syntax for the annotation} \textsc{ }

\lstset{basicstyle=\ttfamily \scriptsize}
\begin{lstlisting}[frame=none]{}
<fuzzyOwl2 fuzzyType="datatype">
  <Datatype type="modified" modifier="<STRING>" base="<STRING>" />
</fuzzyOwl2>
\end{lstlisting}

\paragraph{Semantical restrictions}
The parsers should check the following restrictions:

\begin{itemize}
    \item $modifier$ has already been defined as a fuzzy modifier.
    \item $base$ has already been defined as a fuzzy datatype.
\end{itemize}

\begin{example}
Let us represent the fuzzy datatype $\textsf{VeryYoungAge}$. To
begin with, we assume that the fuzzy datatype $\textsf{YoungAge}$
has been created as in Example~\ref{ex:datatype}, and that the fuzzy
datatype $\textsf{very}$ has been created as in
Example~\ref{ex:modifier}. Next, we define a new datatype
$\textsf{VeryYoungAge}$, adding the following annotation property to
it:

\lstset{basicstyle=\ttfamily \scriptsize}
\begin{lstlisting}[frame=none]{}
<fuzzyOwl2 fuzzyType="datatype">
  <Datatype type="modified" modifier="very" base="YoungAge" />
</fuzzyOwl2>
\end{lstlisting}

\end{example}

\subsection{Fuzzy concepts}

\nd In this case, we create a new concept $D$ and to add an annotation
property describing the type of the constructor and the value of
their parameters. Now, the value of \verb"fuzzyType" is
\verb"concept", and there is a tag \verb"Concept" with an attribute
\verb"type", and other attributes, depending on the concept
constructor. The general rule is that recursion is not allowed,
i.e., $D$ cannot be defined in terms of $D$, so $D$ is not a valid
value for these attributes.

\subsubsection{Fuzzy modified concepts}

\nd Here, the value of \verb"type" is \verb"modified". There are also
two additional attributes: \verb"modifier" (fuzzy modifier), and
\verb"base" (the name of the fuzzy concept that is being modified).

\paragraph{Domain of the annotation}
An OWL 2 concept declaration.

\paragraph{Syntax for the annotation} \textsc{ }

\lstset{basicstyle=\ttfamily \scriptsize}
\begin{lstlisting}[frame=none]{}
<fuzzyOwl2 fuzzyType="concept">
  <MODIFIED_CONCEPT>
</fuzzyOwl2>

<MODIFIED_CONCEPT> := <Concept type="modified" modifier="<STRING>"
base="<STRING>" />
\end{lstlisting}

\paragraph{Semantical restrictions}

The parsers should check the following restrictions:

\begin{itemize}
    \item $modifier$ has already been defined as a fuzzy modifier.
    \item The name of the concept $C$ is different from the name
    of the annotated concept.
\end{itemize}

\begin{example}

Let us represent now the concept $\textsf{very}(\textsf{C})$. We
assume that the fuzzy modifier has been created as in
Example~\ref{ex:modifier}. To that end, we create the atomic concept
\textsf{VeryC} and annotate it:

\lstset{basicstyle=\ttfamily \scriptsize}
\begin{lstlisting}[frame=none]{}
Class ( VeryC Annotation( fuzzyLabel
  <fuzzyOwl2 fuzzyType="concept">
    <Concept type="modified" modifier="very" base="C" />
  </fuzzyOwl2>
) )
\end{lstlisting}

\end{example}

\subsubsection{Weighted concepts}

\nd Here, the value of \verb"type" is \verb"weighted". There are also
two additional attributes: \verb"value" (a real number in $(0,1]$),
and \verb"base" (the name of the fuzzy concept that is being
weighted).

\paragraph{Domain of the annotation}
An OWL 2 concept declaration.

\paragraph{Syntax for the annotation} \textsc{ }

\lstset{basicstyle=\ttfamily \scriptsize}
\begin{lstlisting}[frame=none]{}
<fuzzyOwl2 fuzzyType="concept">
  <WEIGHTED_CONCEPT>
</fuzzyOwl2>

<WEIGHTED_CONCEPT> := <Concept type="weighted" value="<DOUBLE>" base="<STRING>" />
\end{lstlisting}

\paragraph{Semantical restrictions}

The parsers should check the following restrictions:

\begin{itemize}
    \item $value$ in $(0,1]$.
    \item The name of the concept $C$ is different from the name
    of the annotated concept.
\end{itemize}

\begin{example}

Let us represent now the concept $(0.8 \; \textsf{C})$. We create
the atomic \textsf{Weight0.8C} and annotate it:

\lstset{basicstyle=\ttfamily \scriptsize}
\begin{lstlisting}[frame=none]{}
Class ( Weight0.8C Annotation( fuzzyLabel
  <fuzzyOwl2 fuzzyType="concept">
    <Concept type="weighted" value="0.8" base="C" />
  </fuzzyOwl2>
) )
\end{lstlisting}

\end{example}

\subsubsection{Weighted sum concepts}

\nd Here, the value of \verb"type" is \verb"weightedSum". There are also
several additional tags representing weighted concepts.

\paragraph{Domain of the annotation}
An OWL 2 concept declaration.

\paragraph{Syntax for the annotation} \textsc{ }

\lstset{basicstyle=\ttfamily \scriptsize}
\begin{lstlisting}[frame=none]{}
<fuzzyOwl2 fuzzyType="concept">
  <Concept type="weightedSum">
    (<WEIGHTED_CONCEPT>)+
  </Concept>
</fuzzyOwl2>
\end{lstlisting}

\paragraph{Semantical restrictions}

Let $k$ be the number of weighted concepts taking part in the
definition. The parsers should check the following restrictions:

\begin{itemize}
    \item $k \geq 2$.
    \item $\sum_{i=1}^{k} value_k \leq 1$.
    \item The names of the concepts $C_i$ are different from the name
    of the annotated concept.
\end{itemize}

\begin{example}

Let us represent now the concept $(0.8 \; \textsf{A} + 0.2 \;
\textsf{B})$. We create the atomic \textsf{Sum08Aplus02B} and
annotate it:

\lstset{basicstyle=\ttfamily \scriptsize}
\begin{lstlisting}[frame=none]{}
Class ( Sum08Aplus02B Annotation( fuzzyLabel
  <fuzzyOwl2 fuzzyType="concept">
    <Concept type="weightedSum">
      <Concept type="weighted" value="0.8" base="A" />
      <Concept type="weighted" value="0.2" base="B" />
    </Concept>
  </fuzzyOwl2>
) )
\end{lstlisting}

\end{example}

\subsubsection{Fuzzy nominals}

\nd Here, the value of \verb"type" is \verb"nominal". There are also two
additional attributes: \verb"value" (a real number in $(0,1]$), and
\verb"individual" (the name of the individual that is being
weighted).

\paragraph{Domain of the annotation}
An OWL 2 concept declaration.

\paragraph{Syntax for the annotation} \textsc{ }

\lstset{basicstyle=\ttfamily \scriptsize}
\begin{lstlisting}[frame=none]{}
<fuzzyOwl2 fuzzyType="concept">
  <FUZZY_NOMINAL_CONCEPT>
</fuzzyOwl2>

<FUZZY_NOMINAL_CONCEPT> := <Concept type="nominal" value=<DOUBLE> individual=<STRING> />
\end{lstlisting}

\paragraph{Semantical restrictions}

The parsers should check the following restrictions:

\begin{itemize}
    \item $value \in (0,1]$.
\end{itemize}

\begin{example}

Let us represent now the concept $\{ 0.75 / \textsf{ind} \}$. We
create the atomic \textsf{ind075} and annotate it:

\lstset{basicstyle=\ttfamily \scriptsize}
\begin{lstlisting}[frame=none]{}
Class ( ind075 Annotation( fuzzyLabel
  <fuzzyOwl2 fuzzyType="concept">
    <Concept type="nominal" value="0.75" individual="ind" />
  </fuzzyOwl2>
) )
\end{lstlisting}

\end{example}

\subsection{Fuzzy roles}

\nd In this case, we create a new concept $R$ and to add an annotation
property describing the type of the constructor and the value of
their parameters. Now, the value of \verb"fuzzyType" is \verb"role",
and there is a tag \verb"Role" with an attribute \verb"type", and
other attributes, depending on the role constructor. The general
rule is that recursion is not allowed. For the moment, we only support fuzzy modified roles.

\subsubsection{Fuzzy modified roles}

\nd Here, the value of \verb"type" is \verb"modified". There are also
two additional attributes: \verb"modifier" (fuzzy modifier), and
\verb"base" (the name of the fuzzy role that is being modified).

\paragraph{Domain of the annotation}
An OWL 2 (object or data) property declaration.

\paragraph{Syntax for the annotation} \textsc{ }

\lstset{basicstyle=\ttfamily \scriptsize}
\begin{lstlisting}[frame=none]{}
<fuzzyOwl2 fuzzyType="role">
  <MODIFIED_ROLE>
</fuzzyOwl2>

<MODIFIED_ROLE> := <Role type="modified" modifier="<STRING>"
base="<STRING>" />
\end{lstlisting}

\paragraph{Semantical restrictions}

The parsers should check the following restrictions:

\begin{itemize}
    \item $modifier$ has already been defined as a fuzzy modifier.
    \item The name of the role $R$ is different from the name
    of the annotated concept.
\end{itemize}

\begin{example}

Let us represent now the abstract role $\textsf{very}(\textsf{R})$.
We assume that the fuzzy modifier has been created as in
Example~\ref{ex:modifier}. To that end, we create the atomic object
property \textsf{VeryR} and annotate it:

\lstset{basicstyle=\ttfamily \scriptsize}
\begin{lstlisting}[frame=none]{}
ObjectProperty ( VeryR Annotation( fuzzyLabel
  <fuzzyOwl2 fuzzyType="role">
    <Role type="modified" modifier="very" base="R" />
  </fuzzyOwl2>
) )
\end{lstlisting}

\end{example}

\subsection{Fuzzy axioms}

\nd It is possible to add a degree of truth to some axioms, i.e.,
(A1)--(A5), (A8), (A12)--(A13). The value of \verb"fuzzyType" is
\verb"axiom". There is an optional tag \verb"Degree", with and
attribute \verb"value". If omitted, we assume degree $1$.

It would also be possible to specify an inequality sign but we will
assume $\geq$. An axiom of the form $\langle \tau > \alpha \rangle$
is equivalent to $\langle \tau \geq \alpha + \epsilon \rangle$.
Regarding axioms involving $\lhd$, note that $\langle \tau \lhd
\alpha \rangle$ is equivalent to $\langle \tau \lhd^- 1 - \alpha
\rangle$\footnote{$\bowtie^-$ denotes the reflection of the operator
$\bowtie$ and is defined as follows: $\geq^- = \leq, >^- = <,
\leq^-=\geq <^- = >$.} in axioms (A1)--(A5). In axioms $(A8), (A12),
(A13)$ we argue that it does not make sense to have axioms of the
form $\langle \tau \lhd \alpha \rangle$ because such axioms do not
have an equivalent expression in classical DLs.

\paragraph{Domain of the annotation}
An OWL 2 axiom of the following types: concept assertion,role
assertion, GCI, RIA. That is, the crisp equivalents of axioms (A1),
(A1)--(A5), (A8), (A12)--(A13).

\paragraph{Syntax for the annotation} \textsc{ }

\lstset{basicstyle=\ttfamily \scriptsize}
\begin{lstlisting}[frame=none]{}
<fuzzyOwl2 fuzzyType="axiom">
  <Degree value="<DOUBLE>" />
</fuzzyOwl2>
\end{lstlisting}

\paragraph{Semantical restrictions}

The parsers should check the following restrictions:

\begin{itemize}
    \item $value$ in $(0, 1]$.
\end{itemize}

\begin{example}
Let us consider again, in greater detail,
Example~\ref{ex:annotation}. Firstly, we create an OWL 2 concept
assertion:

\lstset{basicstyle=\ttfamily \scriptsize}
\begin{lstlisting}[frame=none]{}
ClassAssertion(paul Tall)
\end{lstlisting}

\nd Then, we annotate it as follows:

\lstset{basicstyle=\ttfamily \scriptsize}
\begin{lstlisting}[frame=none]{}
<fuzzyOwl2 fuzzyType="axiom">
  <Degree value="0.5" />
</fuzzyOwl2>
\end{lstlisting}
\end{example}

\subsection{Ontologies}

\nd We may also annotate the ontology and specify the fuzzy logic to be
considered in the semantics.

The value of \verb"fuzzyType" is \verb"ontology". There is a tag
\verb"FuzzyLogic", with and attribute \verb"logic", that specifies
the default fuzzy logic which is used in the semantics of the fuzzy
ontology.

\paragraph{Domain of the annotation}
An OWL 2 ontology.

\paragraph{Syntax for the annotation} \textsc{ }

\lstset{basicstyle=\ttfamily \scriptsize}
\begin{lstlisting}[frame=none]{}
<fuzzyOwl2 fuzzyType="ontology">
  <FuzzyLogic logic=<LOGIC> />
</fuzzyOwl2>

<LOGIC> := "lukasiewicz" | "zadeh"
\end{lstlisting}

\nd At the moment, we only allow two fuzzy logics, \L ukasiewicz and
Zadeh.

\section{Some Applications of Fuzzy Ontologies}
~\label{sec:examples}

\nd In this section, we will provide some examples illustrating how use
fuzzy ontologies to model the knowledge in real application
problems, and how to encode the fuzzy ontologies using the
methodology explained in Section~\ref{sec:owl2}\footnote{The full
examples may be downloaded from \url{http://www.straccia.info}.}.

\subsection{Matchmaking}

\nd To begin with, we will address the family of matchmaking problems.
The following example is a modified version of the one
in~\cite{BobilloFuzzIEEE2008}.

Assume that a car seller sells a sedan car. A buyer is looking for a
second hand passenger car. Both the buyer as well as the seller have
preferences (restrictions). Our aim is to find the best agreement.
The preferences are as follows. Concerning the buyer:

\begin{enumerate}
    \item If there is an alarm system in the car then he is completely satisfied with paying no more than $22300$,
but he can go up to $22750$ to a lesser degree of satisfaction.
    \item He wants a driver insurance and either a theft insurance or a fire
insurance.
    \item He wants air conditioning and the external color should be either black or grey.
    \item Preferably the price is no more than $22000$, but he can go up to $24000$ to a lesser degree of
satisfaction.
    \item The kilometer warranty is preferably at least 175000, but he may go down to $150000$ to a lesser degree of
satisfaction.
    \item The weights of the preferences $1$--$5$ are $0.1, 0.2, 0.1, 0.2, 0.4$, respectively. The higher the value, the more important the preference is.
    \item There is a strict requirement: he does not want to pay more than $26000$ (buyer reservation value).
\end{enumerate}

\nd Concerning the seller:

\begin{enumerate}

    \item If there is an navigator pack system in the car then he is completely satisfied with a price of at least $22750$,
but he can go down to $22500$ to a lesser degree of satisfaction.
    \item He would prefer to sell the Insurance Plus package.
    \item The kilometer warranty is preferably at most $100000$, but he may go up to $125000$ to a lesser degree of
satisfaction.
    \item The monthly warranty is preferably at most $60$, but he may go up to 72 to a lesser degree of
satisfaction.
    \item If the color is black then the car has air conditioning.
    \item The weights of the preferences $1$--$5$ are, $0.3, 0.1, 0.3, 0.1, 0.2$, respectively. The higher the value, the more important the preference is.
    \item There is a strict requirement: he wants to sell no less than $22000$ (seller reservation value).
\end{enumerate}

\begin{figure}[h]
\begin{center}
\includegraphics[width=0.9 \textwidth]{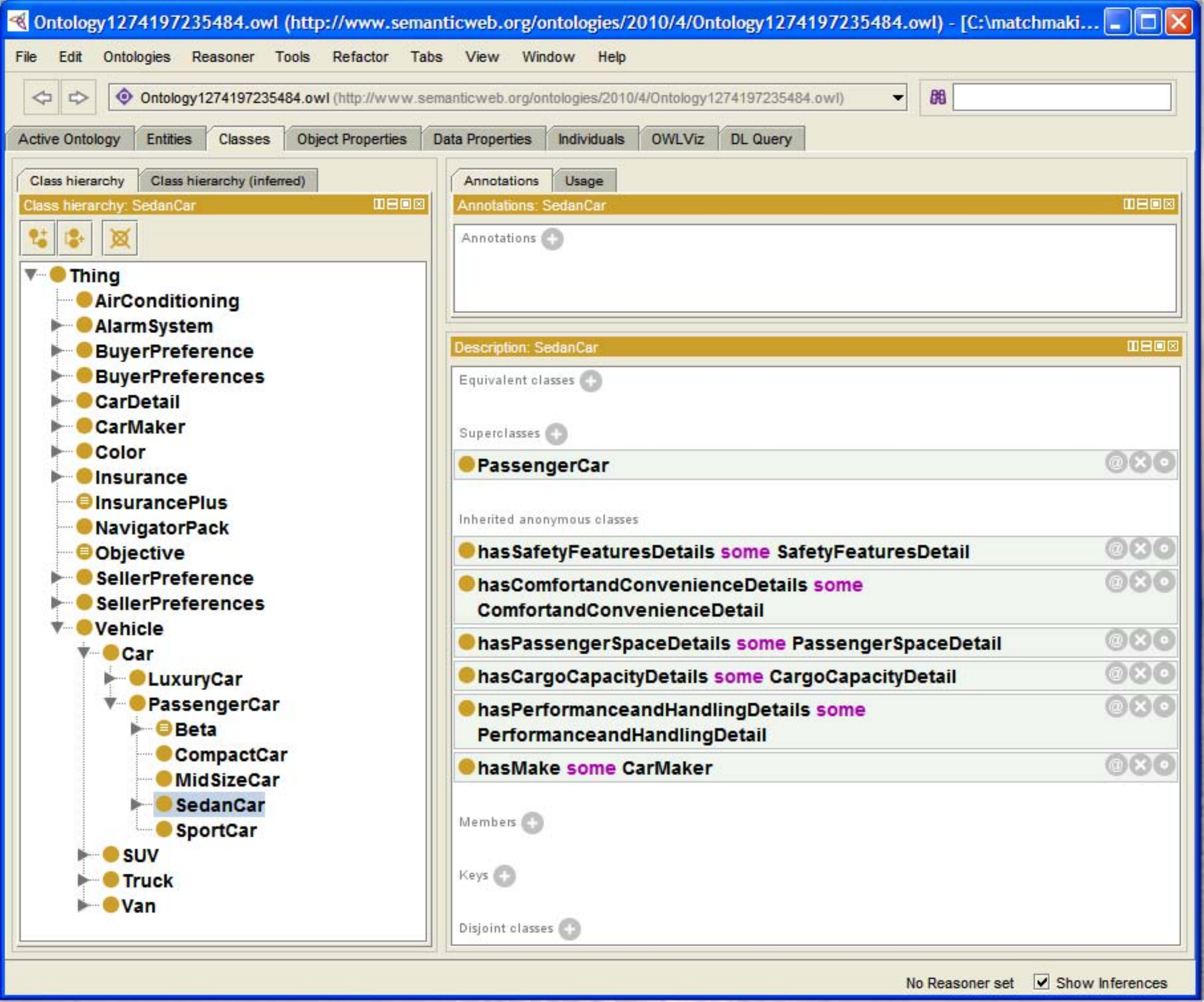}
\caption{Definition of the concept \textsf{Sedan}.}
\label{fig:screenshotSedan}
\end{center}
\end{figure}

\nd We have also some background theory about the domain:

\begin{enumerate}
    \item There are several types of vehicles: car, sport utility vehicle (SUV), truck, and
van. Each of these vehicles has some subclasses. For instance, there
are luxury cars and passenger cars. In particular, a sedan is a
passenger car (see Figure~\ref{fig:screenshotSedan}).
    \item There are several car makers, e.g., BMW, Ferrari,
Volkswagen \dots
    \item There are several car colors, e.g., back, grey \dots
    \item A satellite alarm system is an alarm system.
    \item The \emph{Navigator Pack} is a satellite alarm system with a GPS system.
    \item The \emph{Insurance Plus Package} is a driver insurance together with a theft insurance.
\end{enumerate}

\nd Let us show now how to encode the previous knowledge. A concept
$\textsf{Buy}$ collects all the buyer's preferences together in such
a way that the higher is the maximal degree of satisfiability of
$\textsf{Buy}$, the more the buyer is satisfied.

$$\begin{tabular}{l}
$\textsf{Buy} = \textsf{BuyerRequirements} \sqcap
\textsf{BuyerPreferences}$ \\
$\textsf{BuyerRequirements} = \textsf{PassengerCar} \sqcap \exists
\textsf{hasPrice}.\textsf{leq26000}$ \\
$\textsf{B1} = \neg (\exists
\textsf{hasAlarmSystem}.\textsf{AlarmSystem}) \sqcup \exists
\textsf{hasPrice}.\textsf{ls22300}\textrm{-}\textsf{22750}$ \\
$\textsf{B2} = (\exists
\textsf{hasInsurance}.\textsf{DriverInsurance}) \sqcap \exists
\textsf{hasInsurance}.(\textsf{TheftInsurance} \sqcup
\textsf{FireInsurance})$ \\
$\textsf{B3} = (\exists
\textsf{hasAirConditioning}.\textsf{AirConditioning}) \sqcap \exists
\textsf{HasExColor}.(\textsf{ExColorBlack} \sqcup
\textsf{ExColorGray})$ \\
$\textsf{B4} = \exists \textsf{hasPrice}.\textsf{ls22000}\textrm{-}\textsf{24000}$ \\
$\textsf{B5} = \exists \textsf{hasKMWarranty}.\textsf{rs15000}\textrm{-}\textsf{175000}$ \\
\end{tabular}$$

\nd \textsf{BuyerPreferences} is a weighted sum concept, so we add the
following annotation property to it:

\lstset{basicstyle=\ttfamily \scriptsize}
\begin{lstlisting}[frame=none]{}
<fuzzyOwl2 fuzzyType="concept">
  <Concept type="weightedSum">
    <Concept type="weighted" value="0.1" base="B1" />
    <Concept type="weighted" value="0.2" base="B2" />
    <Concept type="weighted" value="0.1" base="B3" />
    <Concept type="weighted" value="0.2" base="B4" />
    <Concept type="weighted" value="0.4" base="B5" />
  </Concept>
</fuzzyOwl2>
\end{lstlisting}

\begin{figure}[h]
\begin{center}
\includegraphics[width=0.9 \textwidth]{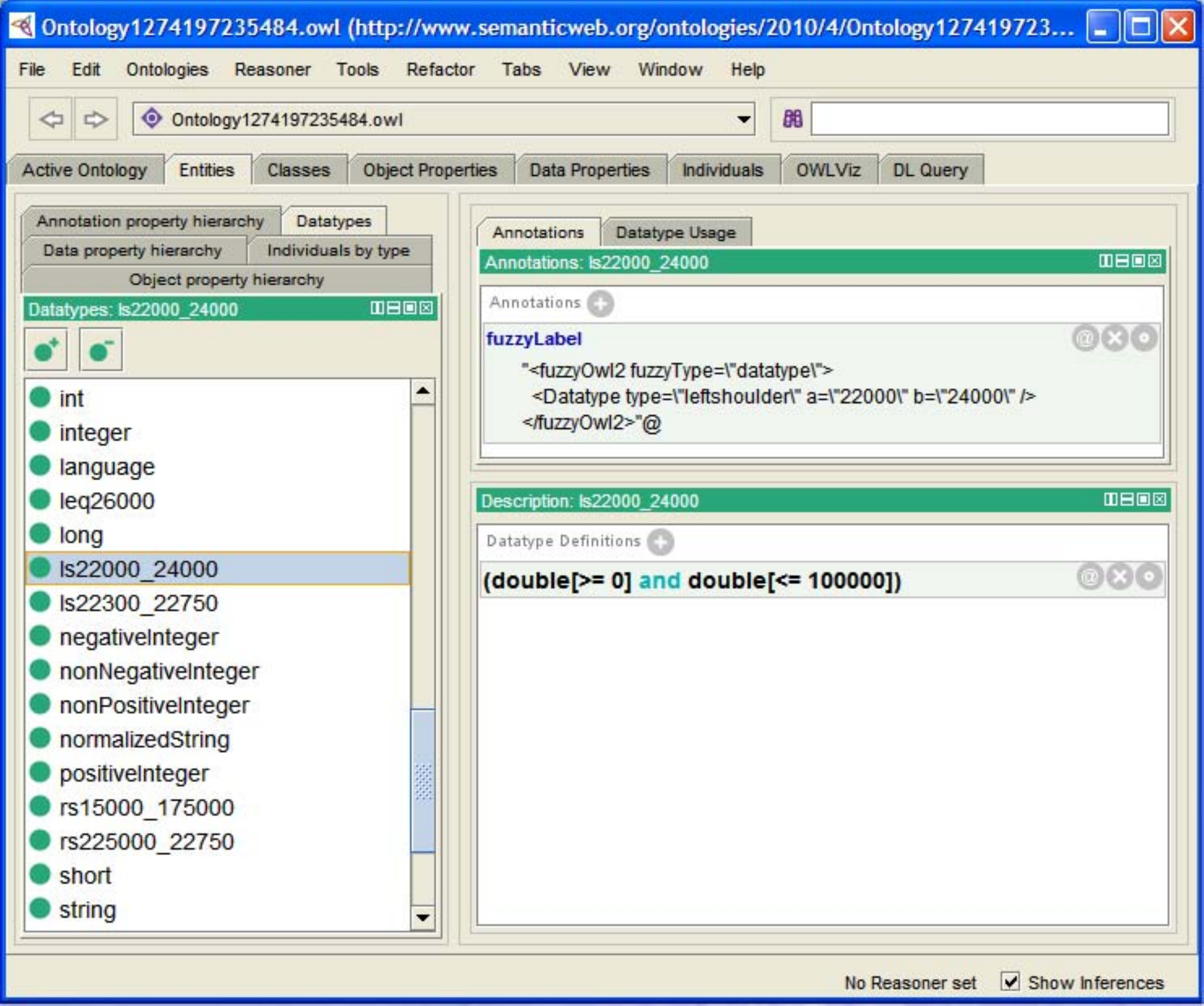}
\caption{Annotation property defining fuzzy datatype
$\textsf{ls22000}\textrm{-}\textsf{24000}$.}
\label{fig:screenshotLs}
\end{center}
\end{figure}

\nd $\textsf{leq26000}$, $\textsf{ls22300}\textrm{-}\textsf{22750}$,
$\textsf{ls22000}\textrm{-}\textsf{24000}$, and
$\textsf{rs15000}\textrm{-}\textsf{175000}$ are defined datatypes
with annotation properties. For instance,
$\textsf{ls22000}\textrm{-}\textsf{24000}$ has the following
annotation property (see Figure~\ref{fig:screenshotLs}):

\lstset{basicstyle=\ttfamily \scriptsize}
\begin{lstlisting}[frame=none]{}
<fuzzyOwl2 fuzzyType="datatype">
  <Datatype type="leftshoulder" a="22000" b="24000" />
</fuzzyOwl2>
\end{lstlisting}

\nd Note that if $a = b$, then we have a crisp concept. This is the case
of the datatype $\textsf{leq26000}$, which is represented as
follows:

\lstset{basicstyle=\ttfamily \scriptsize}
\begin{lstlisting}[frame=none]{}
<fuzzyOwl2 fuzzyType="datatype">
  <Datatype type="leftshoulder" a="26000" b="26000" />
</fuzzyOwl2>
\end{lstlisting}

\nd Similarly to the buyer case, the concept $\textsf{Sell}$ collects
all the seller's preferences together in such a way that the higher
is the maximal degree of satisfiability of $\textsf{Sell}$, the more
the seller is satisfied.

$$\begin{tabular}{l}
$\textsf{Sell} = \textsf{SellerRequirements} \sqcap \textsf{SellerPreferences}$ \\
$\textsf{SellerRequirements} = \textsf{SedanCar} \sqcap \exists
\textsf{hasPrice}.\textsf{geq22000}$ \\
$\textsf{S1} = \neg (\exists
\textsf{hasNavigator}.\textsf{NavigatorPack}) \sqcup \exists
\textsf{hasPrice}.\textsf{rs225000}\textrm{-}\textsf{22750}$ \\
$\textsf{S2} = \exists \textsf{hasInsurance}.\textsf{InsurancePlus}$ \\
$\textsf{S3} = \exists \textsf{hasKMWarranty}.\textsf{SellerKmWarr}$ \\
$\textsf{S4} = \exists \textsf{hasMWarranty}.\textsf{SellerMWarr}$ \\
$\textsf{S5} = \neg (\exists
\textsf{hasExColor}.\textsf{ExColorBlack}) \sqcup \exists
\textsf{hasAirConditioning}.\textsf{AirConditioning}$ \\
\end{tabular}$$

\begin{figure}[ht]
\begin{center}
\includegraphics[width=0.9 \textwidth]{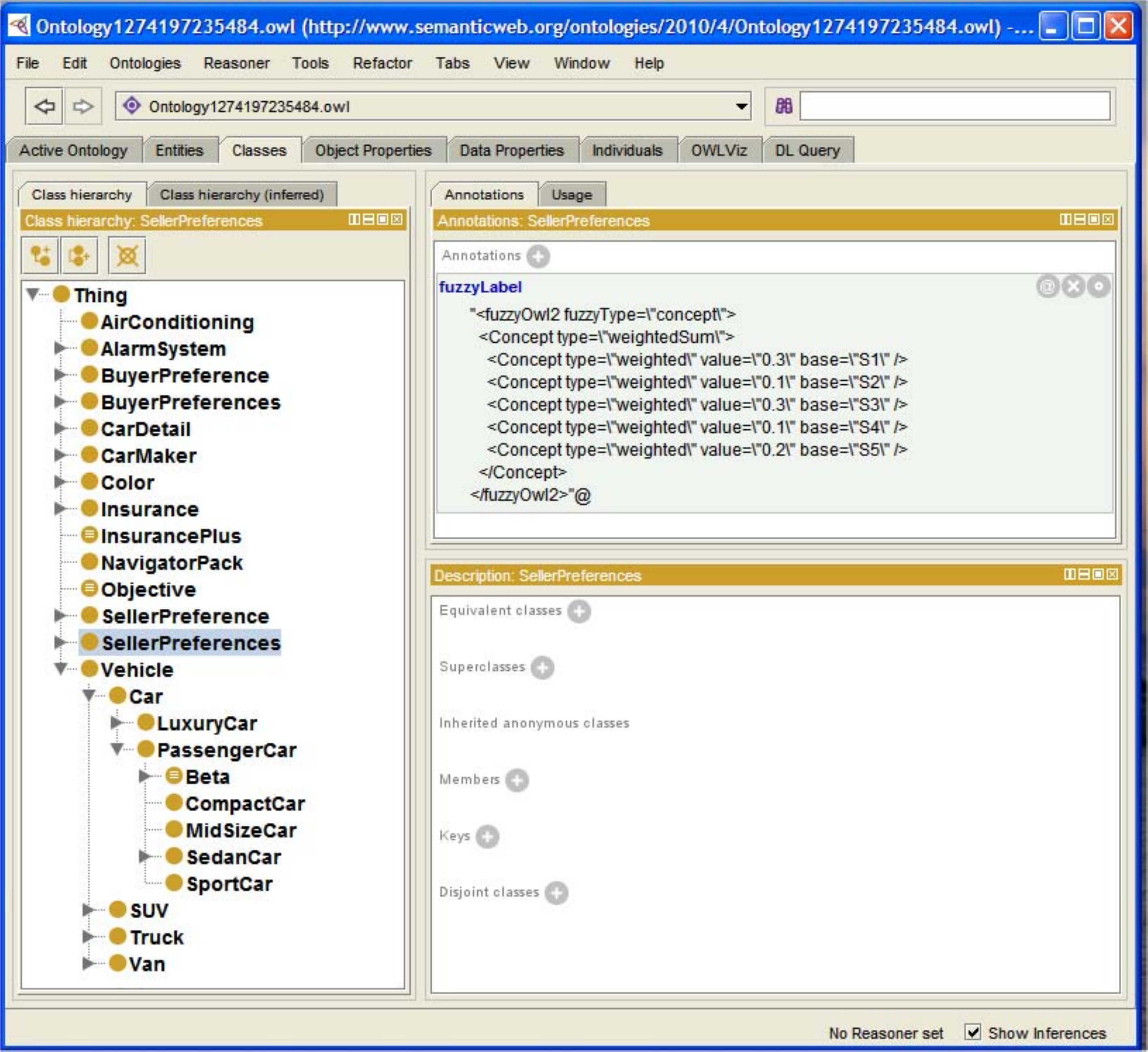}
\caption{Annotation property defining concept
\textsf{SellerPreferences}.} \label{fig:screenshotWsum}
\end{center}
\end{figure}

\nd \textsf{SellerPreferences} is a weighted sum concept, so we add the
following annotation property to it (see
Figure~\ref{fig:screenshotWsum}):

\lstset{basicstyle=\ttfamily \scriptsize}
\begin{lstlisting}[frame=none]{}
<fuzzyOwl2 fuzzyType="concept">
  <Concept type="weightedSum">
    <Concept type="weighted" value="0.3" base="S1" />
    <Concept type="weighted" value="0.1" base="S2" />
    <Concept type="weighted" value="0.3" base="S3" />
    <Concept type="weighted" value="0.1" base="S4" />
    <Concept type="weighted" value="0.2" base="S5" />
  </Concept>
</fuzzyOwl2>
\end{lstlisting}

\nd Similar as in the case of the buyer, \textsf{geq22000},
\textsf{rs225000-22750}, \textsf{SellerKmWarr}, \textsf{SellerMWarr}
are defined datatypes. For instance, \textsf{SellerKmWarr} is
defined as:

\lstset{basicstyle=\ttfamily \scriptsize}
\begin{lstlisting}[frame=none]{}
<fuzzyOwl2 fuzzyType="datatype">
  <Datatype type="leftshoulder" a="100000" b="125000" />
</fuzzyOwl2>
\end{lstlisting}

\nd Now, it is clear that the best agreement among the buyer and the
seller is determined by the maximal degree of satisfiability of the
conjunction $\textsf{Buy} \sqcap \textsf{Sell}$ under \L ukasiewicz
fuzzy logic. So, an optimal match (the degree is $0.7625$) would be
an agreement on a price of $22500$, with $100000$ kilometer warranty
and $60$ month warranty.


\subsection{Multi-criteria Decision Making}

\nd Now, we will concentrate in the family of fuzzy multi-criteria
decision making (MCDM) problems. The following example is a modified
version of the one in~\cite{StracciaKES2010}.

Given a set of $n$ decision alternatives and a set of $m$ criteria
according to which the desirability of an action is judged, a MCDM
problem of $m$ criteria and $n$ alternatives consists in determining
the optimal alternative $a^\star$ with the highest degree of
desirability.

Usually, \emph{alternatives} represent different choices of action
available to the decision maker. The decision \emph{criteria} (also
referred to as goals or attributes) represent the different
dimensions from which the alternatives can be viewed. For instance,
cost, quality, or delivery time. A standard feature of MCDM methods
is that a MCDM problem can be expressed by means of a \emph{decision
matrix}. In the matrix, each row corresponds to an alternative
$a_i$, and each column belongs to a criterion $c_j$. The score
$p_{ij}$ describes the \emph{performance} of alternative $a_i$
against criterion $c_j$.

Most of the MCDM methods require to establish the relative
importance of every criterion in the decision by assigning a
\emph{weight} to it. The weights of the criteria are usually
determined on subjective basis and may also be seen as a kind of
profit of the criteria. Usually, these weights are normalized to add
up to one.

We assume the existence of some \emph{experts} $e_k$ that define the
performances and the weights. Given a criterion $c_j$, the expert
$e_k$ associated to it a relative importance $w^k_j \in [0,1]$ such
that $\sum^n_{j=1} w^k_j = 1$. Also, $e_k$ defines the performance
$p^k_{ij}$ for each alternative $i$ and for each criterion $j$ by
means of a fuzzy number. In fuzzy MCDM, the principal difference
with the classical case is actually the fact that performance
factors are \emph{fuzzy numbers} defined by means of triangular
membership functions $\mathtt{triangular}(a, b, c)$, which are
intended to be an approximation of the number $b$.

For instance, if there are $2$ experts, $2$ alternatives and $2$
criteria, we may have the following decision matrix:

\begin{center}{\small
\begin{tabular}{|c|c|c|}
\hline
$e_1$ & $c_1$                & $c_2$ \\
\hline
$a_1$ & $\mathtt{triangular}(0.6, 0.7, 0.8)$ &
$\mathtt{triangular}(0.9, 0.95, 1)$ \\
\hline
$a_2$ & $\mathtt{triangular}(0.6, 0.7, 0.8)$   &
$\mathtt{triangular}(0.4, 0.5, 0.6)$ \\
\hline \hline
$e_2$ & $c_1$                & $c_2$ \\
\hline
$a_1$ & $\mathtt{triangular}(0.55, 0.6, 7)$  & $\mathtt{triangular}(0.4, 0.45, 0.5)$ \\
\hline
$a_2$ & $\mathtt{triangular}(0.35, 0.4, 0.45)$ & $\mathtt{triangular}(0.5, 0.55, 0.6)$ \\
\hline
\end{tabular}
}\end{center}

\nd For this decision matrix, we may have the following weights $w^k_j$:

\begin{center}{\small
\begin{tabular}{|c|c|c|}
\hline
      & $c_1$  & $c_2$  \\
\hline
$e_1$ & $0.48$ & $0.52$ \\
\hline
$e_2$ & $0.52$ & $0.48$ \\
\hline
\end{tabular}
}\end{center}

\nd There are many alternative methods to compute the final ranking
values from the decision matrix. We will use the \emph{Weighted Sum
Method} (WSM), which is among the simplest methods in MCDM, but has
the advantage to be easy embedded within fuzzy DLs. Formally, $A^k_i
= \sum^m_{j=1} p^k_{ij} w^k_j$ is the the final ranking value of the
alternative $a_i$ according to the expert $k$.

The final ranking value of the alternative $a_i$ is obtained as an
average of the values obtained for every expert: $A_i = \sum^p_{k=1}
A^k_{i}$.

The \emph{ranking of the alternatives} is obtained by ordering the
alternatives in descending order with respect to the final ranking
value and the optimal alternative $a^\star$ is the one that
maximizes the final ranking value, i.e., $a^\star = \arg\max_{a_i}
A_i$.

Let us show now how to encode the previous knowledge.
Every triangular membership function in the decision matrix is
represented using a datatype with an annotation property indicating
the parameters of the triangular membership function. For every
performance $p^k_{ij}$ we have a defined datatype $\textsf{a-ijk}$.
For instance, the datatype $\textsf{a-211}$ contains the parameters
of the triangular function which defines the performance for the
alternative $2$, criterion $1$, and expert $1$:

\lstset{basicstyle=\ttfamily \scriptsize}
\begin{lstlisting}[frame=none]{}
<fuzzyOwl2 fuzzyType="datatype">
  <Datatype type="triangular" a="0.6" b="0.7" c="0.8" />
</fuzzyOwl2>
\end{lstlisting}

\nd For each alternative $a_i$, for each criterion $c_j$, and for each
expert $e_k$, we define a concept \textsf{Performance-ijk}
establishing the relation with the corresponding cell of the
decision matrix. For instance, $\textsf{Performance-211}$ is defined
as:

$$\begin{tabular}{l}
$\textsf{Performance-211} = \exists \textsf{hasScore}.\textsf{a-211}$ \\
\end{tabular}$$

\nd For each alternative $a_i$, and for each expert $e_k$, we define a
concept $\textsf{LocalValue-ik}$, annotated as a weighted sum
concept. For instance, $\textsf{LocalValue-11}$ is annotated as
follows:

\lstset{basicstyle=\ttfamily \scriptsize}
\begin{lstlisting}[frame=none]{}
<fuzzyOwl2 fuzzyType="concept">
  <Concept type="weightedSum">
    <Concept type="weighted" value="0.48" base="Performance-111" />
    <Concept type="weighted" value="0.52" base="Performance-121" />
  </Concept>
</fuzzyOwl2>
\end{lstlisting}

\nd For each alternative $a_i$, we define a concept
$\textsf{GlobalValue-i}$, annotated as a weighted sum concept. For
instance, $\textsf{GlobalValue-1}$ is annotated as follows:

\lstset{basicstyle=\ttfamily \scriptsize}
\begin{lstlisting}[frame=none]{}
<fuzzyOwl2 fuzzyType="concept">
  <Concept type="weightedSum">
    <Concept type="weighted" value="0.5" base="LocalValoration11" />
    <Concept type="weighted" value="0.5" base="LocalValoration12" />
  </Concept>
</fuzzyOwl2>
\end{lstlisting}

\nd Finally, the best one is the alternative $a_i$ maximizing the
satisfiability degree of the fuzzy concept $\textsf{GlobalValue-i}$.
Following our example, the satisfiability degree of
$\textsf{GlobalValue-1}$ is $0.26$, and the satisfiability degree of
$\textsf{GlobalValue-2}$ is $0.32$. Consequently, the optimal
alternative is $a_2$.


\section{Discussion}
~\label{sec:discussion}

\nd This section discusses the implementation status of our approach and
compares it with the related work.

\subsection{Implementation}

\nd This representation of fuzzy ontologies suggests a methodology for
fuzzy ontology development. First, we can build the \emph{core part}
of the ontology by using any ontology editor supporting OWL 2, such
as \emph{Prot\'{e}g\'{e}} 4.1\footnote{\url{http://protege.stanford.edu/}}~\cite{Fact++11,Protege}.
This allows to reason with this
part using standard ontology reasoners. Then, we add the \emph{fuzzy
part} of the ontology by using annotation properties. This can also
be done directly with an OWL 2 ontology editor.

Once the fuzzy ontology has been created, it has to be translated
into the language supported by some fuzzy DL reasoner, so that we
can reason with it. For this purpose, we have developed a template
code for a parser translating from OWL 2 with annotations of type
\verb"fuzzyLabel" into the language supported by some fuzzy DL
reasoner.

This general parser can be adapted to any particular fuzzy DL
reasoner. As illustrative purposes, we have adapted it to the
languages supported by the fuzzy DL reasoners
\texttt{fuzzyDL}\footnote{\url{http://www.straccia.info/software/fuzzyDL/fuzzyDL.html}}~\cite{BobilloFuzzIEEE2008}
and
\texttt{DeLorean}\footnote{\url{http://webdiis.unizar.es/~fbobillo/delorean}}~\cite{BobilloURSW2008}.
The template and the parsers can be freely obtained from the web
pages of \texttt{fuzzyDL} and \texttt{DeLorean}. It is important to
point out that similar parsers for other fuzzy DL reasoners can be
obtained without difficulties. These three parsers are publicly
available on the
web\footnote{\url{http://www.straccia.info/software/FuzzyOWL/}}.

The parsers are based on \emph{OWL API
3}\footnote{\url{http://owlapi.sourceforge.net}}~\cite{OWLAPI3}. OWL
API 3 is a high level Application Programming Interface for working
with OWL 2 ontologies. It is becoming a de-facto standard and many
SW tools already support it. OWL API allows to iterate over the
elements of the ontology in a transparent way. Whenever an element
is supported by the fuzzy DL reasoner, it is mapped into its
internal representation of a fuzzy ontology. The output of the
process is a fuzzy ontology, which can be printed in the standard
output or saved in a text file.

\begin{table}[htbp]
\caption{Fragments of fuzzy OWL 2 supported by the fuzzy DL
reasoners \texttt{fuzzyDL} and \texttt{DeLorean}.}
\label{tab:fOWL2supported}
\begin{center}{\scriptsize
\begin{tabular}{ll}
%
\begin{tabular}{lll}
  \hline
  \textbf{Concept} & \texttt{fuzzyDL} & \texttt{DeLorean} \\
  \hline
  (C1)  & Yes & Yes \\
  (C2)  & Yes & Yes \\
  (C3)  & Yes & Yes \\
  (C4)  & Yes & Yes \\
  (C5)  & Yes & Yes \\
  (C6)  & Yes & Yes \\
  (C7)  & Yes & Yes \\
  (C8)  & Yes & Yes \\
  (C9)  & Yes & Yes \\
  (C10) & Yes & Yes \\
  (C11) & No  & Yes \\
  (C12) & No  & Yes \\
  (C13) & No  & Yes \\
  (C14) & No  & Yes \\
  (C15) & No  & Yes \\
  (C16) & Yes & Yes \\
  (C17) & Yes & Yes \\
  (C18) & Yes & No  \\
  (C19) & Yes & No  \\
  \hline
  \textbf{Role} & \texttt{fuzzyDL} & \texttt{DeLorean} \\
  \hline
  (R1) & Yes & Yes \\
  (R2) & Yes & Yes \\
  (R3) & No  & Yes \\
  (R4) & No  & No  \\
  (R5) & Yes & Yes  \\
  \hline
\end{tabular}
&
\begin{tabular}{lll}
  \hline
  \textbf{Axiom} & \texttt{fuzzyDL} & \texttt{DeLorean} \\
  \hline
  (A1)  & Yes & Yes \\
  (A2)  & Yes & Yes \\
  (A3)  & No  & Yes \\
  (A4)  & Yes & Yes  \\
  (A5)  & Yes & Yes  \\
  (A6)  & No  & Yes \\
  (A7)  & No  & Yes \\
  (A8)  & Yes & Yes \\
  (A9)  & Yes & Yes \\
  (A10) & Yes & Yes \\
  (A11) & Yes & Yes \\
  (A12) & Partial & Yes \\
  (A13) & Yes & Yes \\
  (A14) & Yes & Yes \\
  (A15) & Yes & Yes \\
  (A16) & Yes & Yes \\
  (A17) & Yes & Yes \\
  (A18) & Yes & Yes \\
  (A19) & Yes & Yes \\
  (A20) & No  & Yes \\
  (A21) & No  & Yes \\
  (A22) & Yes & Yes \\
  (A23) & No  & Yes \\
  (A24) & Yes & Yes \\
  (A25) & No  & Yes \\
  \hline
\end{tabular}
\end{tabular}

}
\end{center}
\end{table}

\nd A full reasoning algorithm for the logic presented in
Section~\ref{sec:fuzzySROIQD} is not known yet. Consequently, the
parsers only cover the fragments of fuzzy OWL 2 currently supported
by these reasoners. Table~\ref{tab:fOWL2supported} summarizes the
fragments of fuzzy OWL 2 supported by \texttt{fuzzyDL} and
\texttt{DeLorean}\footnote{We say that \texttt{fuzzyDL} partially
supports axioms (A12) because they are restricted to the case $m =
1$.}. Such table should not be intended as a comparison of the two
reasoners. Even if \texttt{fuzzyDL} is based of fuzzy
$\mathcal{SHIF}\mathbf{(D)}$ instead of fuzzy
$\mathcal{SROIQ}\mathbf{(D)}$, there are many features that are not
available in other fuzzy DL reasoner.

\subsection{Related work}

\nd This is, to the best of our knowledge, the first effort towards
fuzzy ontology representation using OWL 2.

Na\"{\i}ve fuzzy extensions of ontology languages have been
presented, more precisely OWL~\cite{FuzzyOWLChino,StoilosIJAR} and
OWL 2~\cite{StoilosSROIQ}. These languages are obviously not
complaint with OWL 2 and current ontology editors, as it happens
under out approach. Furthermore, they are not expressive enough
since they only allow a fuzzy ABox. That is, they are restricted to
a subset of our case 5, only for axioms (A1)--(A3).

A similar work provides an OWL ontology for fuzzy ontology
representation~\cite{BobilloISMIS2009}. There, annotation properties
are not used, but concepts, roles and axioms are represented as
individuals. For instance, Example~\ref{ex:axioms} would be
represented using the following axioms (in abstract syntax):

\lstset{basicstyle=\ttfamily \scriptsize}
\begin{lstlisting}[frame=none]{}
(ClassAssertion paul Individual)
(ClassAssertion tall Concept)
(ClassAssertion ax1 ConceptAssertion)
(ObjectPropertyAssertion ax1 isComposedOfAbstractIndividual)
(ObjectPropertyAssertion ax1 isComposedOfAbstractConcept)
\end{lstlisting}

\nd However, this representation has many problems:

\begin{itemize}
    \item Representing concepts, roles and axioms as individuals causes
(meta)logical problems.
    \item Instead of reusing current ontology editors, the method
requires a completely different and user-unfriendly way of
modelling, e.g., a concept conjunction is not represented using
\verb"intersectionOf", but using a specific encoding using a
individual (representing the concept) related with two individuals
(each of them representing one of the conjuncts).
    \item Last but no least, it is not an efficient representation, since the
ontology grows exponentially with the size of the ontology.
\end{itemize}

\nd A closer approach to ours is~\cite{ProntoISWC2008}, which  also uses
annotation properties to add probabilistic constraints, but it is
restricted to a subset of our case 5, axioms (A1) and (A8).

A pattern for uncertainty representation in ontologies has also been
presented in~\cite{VacuraTSD}. However, it is restricted to a subset
of our case 5, only for axioms (A1). Furthermore, it relies in OWL
Full, thus not making possible to reason with the ontology.

Our approach should not be confused with a series of works that
describe, given a fuzzy ontology, how to obtain an equivalent OWL 2
ontology (see for
example~\cite{BobilloIJUFKS,BobilloIJAR,BobilloISMIS2008,StoilosSROIQ,StracciaCrispALCH}).
In these works it is possible to reason using a crisp DL reasoner
instead of a fuzzy DL reasoner, which is not our case. However, the
advantage of our approach is that we provide a specific format to
represent fuzzy ontologies which can be easily managed by current
OWL editors and understood by humans.

The W3C Uncertainty Reasoning for the World Wide Web Incubator Group
(URW3-XG) defined an ontology of uncertainty, a vocabulary which can
be used to annotate different pieces of information with different
types of uncertainty (e.g. vagueness, randomness or incompleteness),
the nature of the uncertainty, etc.~\cite{URW3}. But unlike our
approach, it can only be used to identify some kind of uncertainty,
and not to represent and manage uncertain pieces of information.

Finally, we explain the main differences with a previous version of
our work~\cite{BobilloFuzzIEEE2010a}.

\begin{itemize}

\item In the previous version, there are some concept constructors that have several versions
depending on the fuzzy logic considered. For instance, we had
$\sqcap_G$ and $\sqcap_{\L}$ denoting G\"{o}del and \L ukasiewicz
t-norm, respectively. This has the advantage that the user is free
to combine connectives from different fuzzy connectives. However,
this also has some problems. Firstly, from a practical point of
view, such combinations are not clear yet from a reasoning point of
view. Secondly, since OWL 2 does not make possible to annotate
concept expressions, this would require to create a new named entity
every time these constructors are used, which is problematic from a
modelling point of view. For instance, given a concept $C_1 \sqcap_G
C_2$ would require to create a new concept $D = C_1 \sqcap C_2$ and
to annotate it with the semantics of the fuzzy logic.

\item In the previous version, there also some axioms which
have several versions depending on the fuzzy logic considered, but
we do not allow them either for the sake of coherence.

\item As a consequence of the previous differences, now we allow to
annotate ontologies, in order to specify the fuzzy logic considered
in the semantics of all the elements of the ontology.

\item In the previous version, we used annotation properties of type
\verb"rdfs:comment". Obviously, there was not a clear separation
between real comments and fuzzy information. This has been solved by
using a new annotation property \verb"fuzzyLabel".

\item In the current version, we are restricted to \L ukasiewicz and Zadeh fuzzy logics,
which are supported by \texttt{fuzzyDL} or \texttt{DeLorean}.
However, it is trivial to extend the syntax to cover alternative
fuzzy logics, such as G\"{o}del or Product.

\end{itemize}


\section{Conclusions and Future Work}
\label{sec:conclusions}

\nd In this article we have dealt with the problem of fuzzy ontology
representation. Instead of proposing a fuzzy extension of an
ontology language as a candidate to become a standard for fuzzy
ontologies, which is not foreseeable in the next years, we have
proposed a framework represent fuzzy ontologies using current
languages and resources.

To begin with, we have claimed that the current fuzzy extensions are
not expressive enough, and have identified the syntactical
differences that a fuzzy ontology language has to cope with,
grouping them into $5$ different cases. Our work consider a very
general fuzzy extension of the DL $\mathcal{SROIQ}\mathbf{(D)}$,
which is the logical formalism of OWL 2. In fact, our logic is not
restricted to a simple fuzzy ABox, but there are many differences
with respect to the case, such as fuzzy datatypes, fuzzy modifiers
or weighted sum concepts. However, our approach is extensible and
can easily be augmented to support, e.g., alternative fuzzy logics,
modifier functions and fuzzy datatypes.

Then, we have provided a representation using the current standard
language OWL 2, by using annotation properties. A similar approach
cannot be represented in OWL DL as it does not support rich enough
annotation capabilities. This way, we can use OWL 2 editors to
develop fuzzy ontologies. Furthermore, non-fuzzy reasoners applied
over such a fuzzy OWL ontology can discard the fuzzy part, i.e., the
annotations, producing the same results as if they would not exist.

This work suggests a methodology for fuzzy ontology development.
First, we can build the \emph{core part} of the ontology by using
any ontology editor supporting OWL 2. This allows to reason with
this part using standard ontology reasoners. Then, we add the
\emph{fuzzy part} of the ontology by using annotation properties.
This can also be done directly with an OWL 2 ontology editor, even
if some sort of user assistance would be highly appreciated.

In this regard, we have also developed some parsers translating from
OWL 2 with annotations of type \verb"fuzzyLabel" into the languages
supported by some fuzzy DL reasoners. Firstly, we develop a general
parser that can be adapted to any fuzzy DL reasoner. Then, as
illustrative purposes, we adapted it to the languages supported by
the fuzzy DL reasoners \texttt{fuzzyDL} and \texttt{DeLorean}.
Similar parsers for other fuzzy DL reasoners could be easily
obtained.

We are currently developing a graphical interface (a Prot\'{e}g\'{e}
plug-in) to make the encoding of annotation properties transparent
to the user. In future work, we would like to develop similar
parsers for other fuzzy DL reasoners, such as \texttt{Fire}.


\section*{Acknowledgement}

F. Bobillo has been partially funded by the Spanish Ministry of
Science and Technology (project TIN2009-14538-C02-01) and Ministry
of Education (program Jos\'{e} Castillejo, grant JC2009-00337).


\begin{thebibliography}{10}

\bibitem{DLHandbook}
F.~Baader, D.~Calvanese, D.~McGuinness, D.~{Nardi}, and P.~F.
  {{Patel-Schneider}}.
\newblock {\em The Description Logic Handbook: Theory, Implementation, and
  Applications}.
\newblock Cambridge University Press, 2003.

\bibitem{BobilloURSW2008}
F.~Bobillo, M.~Delgado, and J.~G\'{o}mez-Romero.
\newblock {DeLorean}: {A} reasoner for fuzzy {OWL} 1.1.
\newblock In {\em Proceedings of the 4th International Workshop on
  Uncertainty Reasoning for the Semantic Web (URSW 2008)}, volume 423 of CEUR
  Workshop Proceedings, 10 2008.

\bibitem{BobilloIJUFKS}
F.~Bobillo, M.~Delgado, and J.~G\'{o}mez-Romero.
\newblock Crisp representations and reasoning for fuzzy ontologies.
\newblock {\em International Journal of Uncertainty, Fuzziness and
  Knowledge-Based Systems}, 17(4):501--530, 2009.

\bibitem{BobilloIJAR}
F.~Bobillo, M.~Delgado, J.~G\'{o}mez-Romero, and U.~Straccia.
\newblock Fuzzy Description Logics under {G\"{o}del} semantics.
\newblock {\em International Journal of Approximate Reasoning}, 50(3):494--514,
  2009.

\bibitem{BobilloFuzzIEEE2008}
F.~Bobillo and U.~Straccia.
\newblock {fuzzyDL}: An expressive fuzzy Description Logic reasoner.
\newblock In {\em Proceedings of the 17th IEEE International Conference on
  Fuzzy Systems (FUZZ-IEEE 2008)}, pages 923--930. IEEE Computer Society, 6
  2008.

\bibitem{BobilloISMIS2008}
F.~Bobillo and U.~Straccia.
\newblock Towards a crisp representation of fuzzy Description Logics under
  {{\L}ukasiewicz} semantics.
\newblock In {\em Proceedings of the 17th International Symposium on Methodologies for
  Intelligent Systems (ISMIS 2008)}, volume 4994 of {\em Lecture Notes in
  Computer Science}, pages 309--318. Springer-Verlag, 5 2008.

\bibitem{BobilloISMIS2009}
F.~Bobillo and U.~Straccia.
\newblock An {OWL} ontology for fuzzy {OWL} 2.
\newblock In J.~R. et~al., editor, {\em Proceedings of the 18th International
  Symposium on Methodologies for Intelligent Systems (ISMIS 2009)}, volume 5722
  of {\em Lecture Notes in Computer Science}, pages 151--160. Springer-Verlag,
  9 2009.

\bibitem{BobilloFuzzIEEE2010a}
F.~Bobillo and U.~Straccia.
\newblock Representing fuzzy ontologies in OWL 2.
\newblock In {\em Proceedings of the 19th IEEE International Conference on
  Fuzzy Systems (FUZZ-IEEE 2010)}, pages 2695--2700. IEEE Press, July 2010.

\bibitem{OWL2JWS}
B.~{Cuenca-Grau}, I.~Horrocks, B.~Motik, B.~Parsia, P.~F.
{Patel-Schneider},
  and U.~Sattler.
\newblock {OWL} 2: {T}he next step for {OWL}.
\newblock {\em Journal of Web Semantics}, 6(4):309--322, 2008.

\bibitem{FuzzyOWLChino}
M.~Gao and C.~Liu.
\newblock Extending {OWL} by fuzzy Description Logic.
\newblock In {\em Proceedings of the 17th IEEE International Conference on
  Tools with Artificial Intelligence (ICTAI 2005)}, pages 562--567. IEEE
  Computer Society, 11 2005.

\bibitem{Hajek98}
P.~H\'{a}jek.
\newblock {\em Metamathematics of Fuzzy Logic}.
\newblock Kluwer, 1998.

\bibitem{OWLAPI3}
M.~Horridge and S.~Bechhofer.
\newblock The {OWL API}: {A} {J}ava {API} for working with owl 2 ontologies.
\newblock In {\em Proceedings of the 6th International Workshop on {OWL}:
  Experiences and Directions (OWLED 2009)}, volume 529 of CEUR Workshop
  Proceedings, 10 2009.

\bibitem{Fact++11}
M.~Horridge, D.~Tsarkov, and T.~Redmond.
\newblock Supporting early adoption of {OWL} 1.1 with {Protege-OWL} and
  {FaCT++}.
\newblock In {\em Proceedings of the 2nd International Workshop on {OWL}: Experience and
  Directions (OWLED 2006)}, volume 216 of CEUR Workshop Proceedings, 11 2006.

\bibitem{ProntoISWC2008}
P.~Klinov and B.~Parsia.
\newblock Optimization and evaluation of reasoning in probabilistic description
  logic: {Towards} a systematic approach.
\newblock In {\em Proceedings of the 7th International
  Semantic Web Conference (ISWC 2008)}, volume 5318 of {\em Lecture Notes in
  Computer Science}, pages 213--228. Springer-Verlag, 10 2008.

\bibitem{StracciaSurvey}
T.~Lukasiewicz and U.~Straccia.
\newblock Managing uncertainty and vagueness in Description Logics for the
  semantic web.
\newblock {\em Journal of Web Semantics}, 6(4):291--308, 2008.

\bibitem{MostertShieldsTheorem}
P.~S. Mostert and A.~L. Shields.
\newblock On the structure of semigroups on a compact manifold with boundary.
\newblock {\em Annals of Mathematics}, 65(1):117--143, 1957.

\bibitem{OWL2syntax}
B.~Motik, P.~F. {Patel-Schneider}, and B.~P. (editors).
\newblock {OWL 2 Web Ontology Language} structural specification and
  functional-style syntax, 2009.
\newblock [Online] Available: \url{http://www.w3.org/TR/owl2-syntax/}.

\bibitem{Protege}
N.~F. Noy, M.~Sintek, S.~Decker, M.~Crubezy, R.~W. Fergerson, and
M.~A. Musen.
\newblock Creating {Semantic Web} contents with {Prot\'{e}g\'{e}-2000}.
\newblock {\em IEEE Intelligent Systems}, 16(2):60--71, 2001.

\bibitem{FLandSW}
E.~Sanchez, editor.
\newblock {\em Fuzzy Logic and the Semantic Web}, volume~1 of {\em Capturing
  Intelligence}.
\newblock Elsevier Science, 2006.

\bibitem{StoilosFire}
G.~Stoilos, N.~Simou, G.~Stamou, and S.~Kollias.
\newblock Uncertainty and the semantic web.
\newblock {\em IEEE Intelligent Systems}, 21(5):84--87, 2006.

\bibitem{StoilosSROIQ}
G.~Stoilos and G.~Stamou.
\newblock Extending fuzzy Description Logics for the semantic web.
\newblock In {\em Proceedings of the 3rd International Workshop on {OWL}:
  Experiences and Directions (OWLED 2007)}, volume 258 of CEUR Workshop
  Proceedings, 6 2007.

\bibitem{StoilosIJAR}
G.~Stoilos, G.~Stamou, and J.~Z. Pan.
\newblock Fuzzy extensions of {OWL}: Logical properties and reduction to fuzzy
  Description Logics.
\newblock {\em International Journal of Approximate Reasoning}, 51:656–--679,
  2010.

\bibitem{Straccia01}
U.~Straccia.
\newblock Reasoning within fuzzy Description Logics.
\newblock {\em Journal of Artificial Intelligence Research}, 14:137--166, 2001.

\bibitem{StracciaCrispALCH}
U.~Straccia.
\newblock Transforming fuzzy Description Logics into classical description
  logics.
\newblock In {\em Proceedings of the
  9th European Conference on Logics in Artificial Intelligence (JELIA 2004)},
  volume 3229 of {\em Lecture Notes in Computer Science}, pages 385--399.
  Springer-Verlag, 9 2004.

\bibitem{StracciaALCD1}
U.~Straccia.
\newblock Description logics with fuzzy concrete domains.
\newblock In {\em Proceedings of the 21st
  Conference on Uncertainty in Artificial Intelligence (UAI 2005)}. AUAI Press,
  7 2005.

\bibitem{StracciaSHOIND}
U.~Straccia.
\newblock A fuzzy Description Logic for the semantic web.
\newblock In E.~Sanchez, editor, {\em Fuzzy Logic and the Semantic Web},
  volume~1 of {\em Capturing Intelligence}, pages 73--90. Elsevier Science,
  2006.

\bibitem{StracciaKES2010}
U.~Straccia.
\newblock Multi-criteria decision making in fuzzy Description Logics: {A} first
  step.
\newblock In {\em Proceedings of the 13th International Conference on
  Knowledge-Based \& Intelligent Information \& Engineering Systems (KES
  2009)}, volume 5711 of {\em Lecture Notes in Artificial Intelligence}, pages
  79--87. Springer-Verlag, 2009.

\bibitem{VacuraTSD}
M.~Vacura, V.~Sv\'{a}tek, and P.~Smr\v{z}.
\newblock A pattern-based framework for representation of uncertainty in
  ontologies.
\newblock In {\em Proceedings of the 11th International Conference on Text, Speech, and
  Dialogue (TSD 2008)}, volume 5246 of {\em Lecture Notes in Computer Science},
  pages 227--234. Springer-Verlag, 9 2008.

\bibitem{URW3}
{W3C Incubator Group on Uncertainty Reasoning for the World Wide Web
Final Report}:
\newblock  \url{http://www.w3.org/2005/Incubator/urw3/XGR-urw3}, 2008.

\bibitem{Zadeh65}
L.~A. Zadeh.
\newblock Fuzzy sets.
\newblock {\em Information and Control}, 8:338--353, 1965.

\end{thebibliography}


\appendix

\section{From Fuzzy DLs to Fuzzy OWL}

\nd In this article we have used fuzzy DLs as the original language to
express fuzzy ontologies. As already claimed throughout this
article, our objective is not provide a new fuzzy ontology language,
such as fuzzy OWL 2. However, for the sake of completeness, we find
useful to include as an appendix, a short note about the relation
between DLs and OWL 2.

An OWL 2 ontology contains descriptions of \emph{classes} (or
concepts in DL terminology), \emph{properties} (roles in DL
terminology) and \emph{individuals}. There are two types of
properties: \emph{object properties} (abstract roles) and
\emph{datatype properties} (concrete roles).
Table~\ref{tab:owl2Concepts} includes the classes and properties
constructors of OWL 2, together with their correspondences in
$\mathcal{SROIQ}\mathbf{(D)}$.

There are two additional types of properties which do not have a
counterpart in the DL, namely \emph{annotation properties}
(\texttt{owl:Annotation\-Property}) and \emph{ontology properties}
(\texttt{owl:Ontology\-Property}), but they just include some
meta-properties of the ontology.

\begin{table}[htbp]
\caption{Class and property constructors in OWL 2}
\label{tab:owl2Concepts} \hbox to \textwidth{ \hss
$$\begin{tabular}{|ll|}
    \hline
    \textbf{OWL 2 abstract syntax} & \textbf{DL syntax} \\
    \hline
    \hline
    Class ($A$) & $A$ \\
    Class (owl:Thing) & $\top$ \\
    Class (owl:Nothing) & $\bot$ \\
    ObjectIntersectionOf ($C,D$) & $C \sqcap D$ \\
    ObjectUnionOf ($C,D$) & $C \sqcup D$ \\
    ObjectComplementOf ($C$) & $\neg C$ \\
    ObjectAllValuesFrom ($R,C$) & $\forall R.C$ \\
    ObjectSomeValuesFrom ($R,C$) & $\exists R.C$ \\
    ObjectHasValue ($R,o$) & $\exists R.\{ o \}$ \\
    DataAllValuesFrom ($T,d$) & $\forall T.\mathbf{d}$ \\
    DataSomeValuesFrom ($T,d$) & $\exists T.\mathbf{d}$ \\
    DataHasValue ($T,v$) & $\exists T.\{ v \}$ \\
    ObjectOneOf ($o_1,\dots,o_m$) & $\{o_1\} \sqcup \{o_2\} \sqcup \{ o_m \}$ \\
    ObjectMinCardinality ($n,S,C$) & $(\geq n\;S.C)$ \\
    ObjectMaxCardinality ($n,S,C$) & $(\leq n\;S.C)$ \\
    ObjectExactCardinality ($n,S,C$) & $(\geq n\;S.C) \sqcap (\leq n\;S.C)$ \\
    ObjectMinCardinality ($n,S$) & $(\geq n\;S.\top)$ \\
    ObjectMaxCardinality ($n,S$) & $(\leq n\;S.\top)$ \\
    ObjectExactCardinality ($n,S$) & $(\geq n\;S.\top) \sqcap (\leq n\;S.\top)$ \\
    DataMinCardinality ($n,T,d$) & $(\geq n\;T.\mathbf{d})$ \\
    DataMaxCardinality ($n,T,d$) & $(\leq n\;T.\mathbf{d})$ \\
    DataExactCardinality ($n,T,d$) & $(\geq n\;T.\mathbf{d}) \sqcap (\leq n\;T.\mathbf{d})$ \\
    DataMinCardinality ($n,T$) & $(\geq n\;T.\top)$ \\
    DataMaxCardinality ($n,T$) & $(\leq n\;T.\top)$ \\
    DataExactCardinality ($n,T$) & $(\geq n\;T.\top) \sqcap (\leq n\;T.\top)$ \\
    ObjectExistsSelf ($S$) & $\exists S.\mathtt{Self}$ \\
    \hline
    \hline
    ObjectProperty ($R_A$) & $R_A$ \\
    TopObjectProperty & $U$ \\
    BottomObjectProperty & $\neg U$ \\
    DatatypeProperty ($T$) & $T$ \\
    TopDataProperty & $U_{D}$ \\
    BottomDataProperty & $\neg U_{D}$ \\
    \hline
\end{tabular}$$
\hss }
\end{table}

An OWL 2 document consists of optional ontology headers plus any
number of axioms: facts about individuals, class axioms and property
axioms, which according to the DL terminology correspond to the
ABox, TBox and RBox, respectively. Ontology headers are used for
meta-information, ontology import and relationships.
Table~\ref{tab:owl2Axioms} shows the OWL 2 axioms and their
equivalences in $\mathcal{SROIQ}\mathbf{(D)}$.

\begin{table}[htbp]
\caption{Axioms in OWL 2} \label{tab:owl2Axioms} \hbox to
\textwidth{ \hss

$$\begin{tabular}{|ll|}
    \hline
    \textbf{OWL 2 abstract syntax} & \textbf{DL syntax} \\
    \hline \hline
    ClassAssertion ($a,C$) & $a \colon C$ \\

    ObjectPropertyAssertion ($R,a,b$)) & $(a,b) \colon R$ \\

    NegativeObjectPropertyAssertion ($R,a,b$) & $(a,b) \colon \neg R$ \\

    DataPropertyAssertion ($T,a,v$)) & $(a,v) \colon T$ \\

    NegativeDataPropertyAssertion ($T,a,v$) & $(a,v) \colon \neg T$ \\

    SameIndividual ($a_1, \dots, a_m$) & $a_i = a_j, 1 \leq i < j \leq m$ \\

    DifferentIndividuals ($a_1, \dots, a_m$) & $a_i \neq a_j, 1 \leq i < j \leq m$ \\

    SubClassOf ($C_1,C_2$) & $C_1 \sqsubseteq C_2$ \\

    EquivalentClasses ($C_1,\dots,C_m$) & $C_1 \equiv \dots \equiv C_m$ \\

    DisjointClasses ($C_1,\dots,C_m$) & $\texttt{dis}(C_{1}, \dots, C_{m})$ \\

    DisjointUnion ($C,C_1,\dots,C_m$) & $\texttt{disUnion}(C_{1}, \dots, C_{m})$ \\

    SubObjectPropertyOf (subObjectPropertyChain ($R_1, \dots, R_m$) $R$) & $R_{1} \dots R_{m} \sqsubseteq R$ \\

    SubObjectPropertyOf ($R_1 R_2$) & $R_{1} \sqsubseteq R_2$ \\

    SubDataPropertyOf ($T_1,T_2$)  &  $T_{1} \sqsubseteq T_2$ \\

    EquivalentObjectProperties ($R_1,\dots,R_m$) & $R_1 \equiv \dots \equiv R_m$ \\

    EquivalentDataProperties ($T_1,\dots,T_m$) & $T_1 \equiv \dots \equiv T_m$ \\

    ObjectPropertyDomain ($R,C$) & $\texttt{domain}(R,C)$ \\

    ObjectPropertyRange ($R,C$) & $\texttt{range}(R,C)$ \\

    DataPropertyDomain ($T,d$) & $\texttt{domain}(T,\mathbf{d})$ \\

    DataPropertyRange ($T,d$) & $\texttt{range}(T,\mathbf{d})$ \\

    InverseObjectProperties ($R_1,R_2$) & $R_1 \equiv R_2^{-}$  \\

    FunctionalObjectProperty ($S$) & $\texttt{func}(S)$ \\

    FunctionalDataProperty ($T$) & $\texttt{func}(T)$ \\

    InverseFunctionalObjectProperty ($S$) & $\texttt{func}(S^-)$ \\

    TransitiveObjectProperty ($R$) & $\texttt{trans}(R)$ \\

    DisjointObjectProperties ($S_1,S_2$) & $\texttt{dis}(S_{1}, S_{2})$ \\

    DisjointDataProperties ($T_1,T_2$) & $\texttt{dis}(T_{1}, T_{2})$ \\

    ReflexiveObjectProperty ($R$) & $\texttt{ref}(R)$ \\

    IrreflexiveObjectProperty ($S$) & $\texttt{irr}(S)$ \\

    SymmetricObjectProperty ($R$) & $\texttt{sym}(R)$ \\

    AsymmetricObjectProperty ($S$) & $\texttt{asy}(S)$ \\
    \hline
\end{tabular}$$

\hss }
\end{table}

\end{document}